\documentclass[conference]{IEEEtran}
\IEEEoverridecommandlockouts
\usepackage{times}  % DO NOT CHANGE THIS
\usepackage{helvet}  % DO NOT CHANGE THIS
\usepackage{courier}  % DO NOT CHANGE THIS
\usepackage[hyphens]{url}  % DO NOT CHANGE THIS
\usepackage{graphicx} % DO NOT CHANGE THIS
\usepackage{algorithm}
\usepackage{algorithmic}

\usepackage{booktabs}
\usepackage{multirow}
\usepackage[table,xcdraw]{xcolor}
\usepackage{newfloat}
\usepackage{listings}
\begin{document}

\title{HELIOS: Hierarchical Graph Abstraction for Structure-Aware LLM Decompilation}

\author{
    \IEEEauthorblockN{Yonatan~Gizachew~Achamyeleh,  Harsh~Thomare,   
    Mohammad~Abdullah~Al~Faruque}
    {\textit{Dept. of Electrical Engineering and Computer Science, University of California, Irvine, CA, USA}}
    \\
    {\textit{\{yachamye, hthomare, alfaruqu\}@uci.edu}}
}
% \and
% \and
% \and
% \and
% \and

\maketitle

\begin{abstract}
Large language models (LLMs) have recently been applied to binary decompilation, yet they still treat code as plain text and ignore the graphs that govern program control flow. This limitation often yields syntactically fragile and logically inconsistent output, especially for optimized binaries. This paper presents \textsc{HELIOS}, a framework that reframes LLM based decompilation as a structured reasoning task. \textsc{HELIOS} summarizes a binary's control flow and function calls into a hierarchical text representation that spells out basic blocks, their successors, and high level patterns such as loops and conditionals. This representation is supplied to a general purpose LLM together with raw decompiler output, optionally combined with a compiler in the loop that returns error messages when the generated code fails to build. 

On HumanEval-Decompile for \texttt{x86\_64}, \textsc{HELIOS} raises average object file compilability from 45.0\% to 85.2\% for Gemini~2.0 and from 71.4\% to 89.6\% for GPT\mbox{-}4.1~Mini. With compiler feedback, compilability exceeds 94\% and functional correctness improves by up to 5.6 percentage points over text only prompting. Across six architectures drawn from x86, ARM, and MIPS, \textsc{HELIOS} reduces the spread in functional correctness while keeping syntactic correctness consistently high, all without fine tuning. These properties make \textsc{HELIOS} a practical building block for reverse engineering workflows in security settings where analysts need recompilable, semantically faithful code across diverse hardware targets.

\end{abstract}


\section{Introduction}

Reverse engineering of binary code is a foundational, yet demanding, activity in software security. It underpins tasks such as malware analysis, vulnerability triage, interoperability, and maintenance of legacy systems~\cite{votipka2020observational, lmpa2023}. This work remains a persistent bottleneck, since it requires specialists to reconstruct high level logic from low level, often obfuscated machine instructions. As software complexity and hardware diversity grow~\cite{hennessy2019new}, the demand for skilled reverse engineers exceeds the available expertise~\cite{isaca2021state, darpa_harden, darpa_vspell, achamyeleh2025agnomin}.

The community has explored two main directions for automation. Classical decompilers, such as Ghidra~\cite{ghidra} and IDA Pro~\cite{hexrays}, translate assembly into C like pseudo code and are now standard tools in reverse engineering practice. Their output, however, is often syntactically fragile, poorly typed, and preserves convoluted control flow that still requires substantial manual effort to clean up~\cite{wong2023refining, liu2020how, basque2024ahoy}. More recently, large language models (LLMs) have been used to push past these limitations. Current work either fine tunes LLMs directly on pairs of binaries and source code~\cite{tan2024llm4decompile, jiang2023nova} or prompts general purpose models to rewrite decompiler output~\cite{hu2024degpt, feng_self-constructed_2024, feng_ref_2025}. These approaches differ in training strategy, but they share a common assumption: the binary and its decompiled form can be treated as a one dimensional sequence of tokens. We refer to this family as \textit{structurally blind} (or structurally agnostic) reasoning, since the model receives little or no explicit information about the control flow graph that actually governs program behavior.

\begin{figure*}[t]
    \centering
    \includegraphics[width=0.96\textwidth]{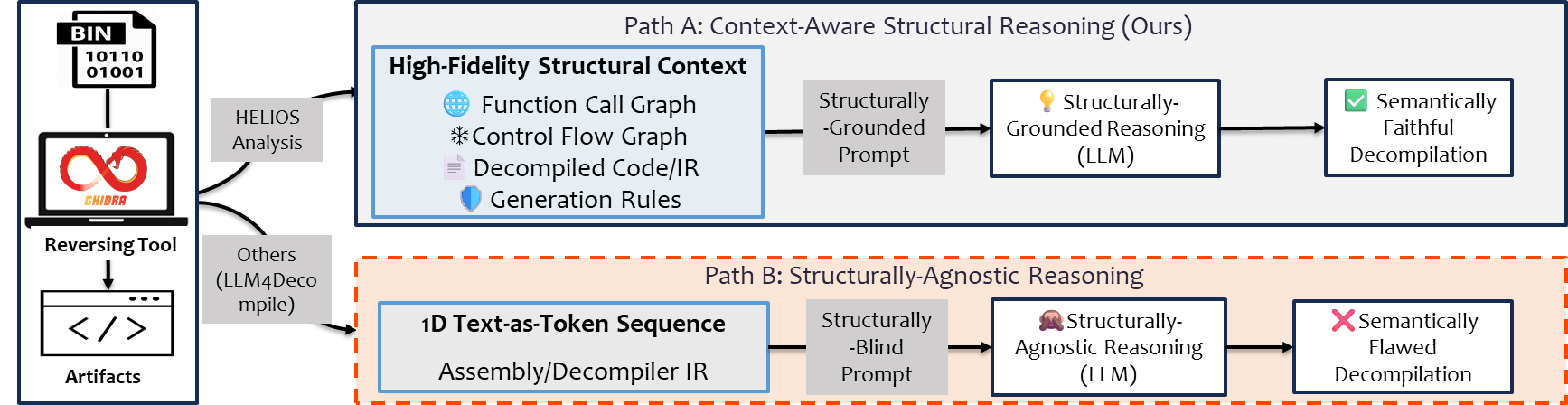}
    \caption{An illustration comparing two approaches for LLM based decompilation. \textbf{Path B (Structurally-Agnostic)}, used by existing methods, treats binary artifacts as flat text. \textbf{Path A (Context-Aware)}, our proposed approach, creates a structural context that allows the LLM to reason over control flow and produce semantically faithful decompilation.}
    \label{fig:conceptual_comparison}
\end{figure*}

Studies of how expert reverse engineers work highlight what this structurally blind view is missing. Observational work~\cite{votipka2020observational, mantovani2022remind} shows that analysts do not simply read assembly in a linear fashion. Instead, they build and refine a mental model of the program by reasoning over graphs, primarily the control flow graph (CFG), supported by the function call graph (FCG) and data flow information~\cite{pennington_stimulus_1987}. This matches a long standing observation in program analysis: graph based representations are a natural abstraction for software, and they are widely used for security tasks such as automated bug finding, function prediction and fuzzing~\cite{yamaguchi2012driller, cha2012unleashing, pei2020trex, cfg2vec, achamyeleh2025agnomin}.

At the same time, current LLMs are text native. They are trained to predict token sequences rather than to traverse graphs. This tension raises a natural question: can we bring the structural view that human analysts use into the token space where LLMs operate, without retraining the model itself. There has been progress on general graph to text encodings~\cite{zhao2023graphtext, fatemi2024talk, guo2023gpt4graph}, but these methods focus on abstract graphs and do not address the specific structure of binaries, such as compiler level intermediate forms, low level control flow, and architecture specific idioms.

This paper introduces \textsc{Helios}, a framework that adapts these ideas to the domain of binary decompilation. We reframe LLM assisted decompilation as context aware structural reasoning (Path A in Figure~\ref{fig:conceptual_comparison}). \textsc{Helios} uses a static analysis backend to derive control flow and call graphs, then encodes this information into a hierarchical textual representation. At a high level, the prompt contains (i) a summary of the function and its role, (ii) a compact description of the CFG and its main paths, and (iii) block level details that link individual basic blocks back to the raw decompiler output. A small set of natural language rules guides the model on how to use this structure, and an optional compiler in the loop provides error messages when the generated code fails to compile. The result is a fine tuning free, architecture agnostic pipeline that encourages the model to use control flow information in a way that is closer to how human analysts work.

In this work, we make the following contributions:
\begin{itemize}
    \item We propose a method for encoding the control flow and call graphs of a binary, together with high level structural patterns such as loops and conditionals, into a compact textual format that can be consumed directly by general purpose LLMs.
    \item We instantiate this method in \textsc{Helios} and evaluate it on HumanEval-Decompile and MBPP, showing large gains in compilability and consistent improvements in functional correctness over text only prompting and over state of the art fine tuned decompilers, while using only a small fixed number of model calls and one optional feedback iteration.
    \item We conduct, to our knowledge, the first cross architecture study of structure aware LLM decompilation across six instruction sets, demonstrating that a single prompt design can generalize across x86, ARM, and MIPS without retraining, which is directly relevant for settings such as firmware and IoT analysis.
\end{itemize}

\section{Background and Related Work}
\label{sec:background}

Our work connects three areas: the representation of binary programs for analysis, the evolution of automated decompilation techniques, and the emerging use of LLMs for reasoning over graph-structured data.

\subsection{Program Representation for Binary Analysis}

Graph-based structures are a standard way to represent the logic of a binary program. The Control Flow Graph (CFG), which models possible execution paths within a function, and the Function Call Graph (FCG), which captures relationships between functions, are widely used in program analysis~\cite{allen1970control, ryder1979constructing, cfg2vec, achamyeleh2025agnomin}. In security, these graph structures are the main data structures for tasks such as automated vulnerability discovery and fuzzing~\cite{feng2016genius, xu2017neural, pei2020trex}. A large body of work has shown that reasoning over these graphs is central to deep program understanding and to discovering subtle bugs~\cite{yamaguchi2012driller, cha2012unleashing}.

\subsection{Automated Decompilation}

The goal of decompilation is to translate low-level machine code back into high-level source code. This is an ill-posed problem, since compilation is inherently lossy~\cite{eilam2011reversing}. Traditional, rule-based decompilers such as Ghidra apply complex heuristics but often produce brittle and semantically incomplete code, for example, by emitting poorly typed variables or misrepresenting control flow~\cite{liu2020how, burk2022decomperson}. As a result, human analysts still spend substantial effort correcting and simplifying decompiler output.

\textit{LLM-based decompilation.}
The emergence of large language models has led to two main lines of work on automated decompilation. The first, which we refer to as \emph{end-to-end decompilation}, uses large-scale fine-tuning. Systems such as LLM4Decompile~\cite{tan2024llm4decompile}, SLaDe~\cite{slade2024}, and Nova~\cite{jiang2023nova} train models on paired source and assembly corpora to translate directly from binaries to high-level code. These approaches can produce high-quality results on the architectures and optimization settings they are trained on, but they require substantial resources and are typically tied to specific instruction sets such as x86\_64. Adapting them to new architectures or to new base models usually requires another round of training.

The second line of work focuses on \emph{decompiler output refinement}. Frameworks such as DeGPT~\cite{hu2024degpt} and LMPA~\cite{lmpa2023}, as well as follow-up work~\cite{wong2023refining}, prompt general-purpose LLMs to clean up or explain pseudo-code produced by existing decompilers. These tools are primarily designed to improve readability and analyst productivity rather than to generate fully recompilable code in a fully automated loop. They mostly operate on the raw decompiler text, with limited access to the underlying control flow or intermediate representations.

Both families of approaches treat the binary and its decompiled form as flattened token sequences. In this sense they are structurally blind, or structurally agnostic, because they do not expose the graph structure that program analysis tools use internally. In contrast, our work keeps the decompiler in the loop. It feeds a general-purpose LLM an explicit, multi-level description of the CFG and related context, aiming to improve recompilability and functional correctness across architectures without additional fine-tuning.

\subsection{Graph Reasoning with Large Language Models}

The mismatch between text native LLMs and graph-structured data has led to a growing literature on methods that linearize graphs into text that an LLM can process~\cite{zhao2023graphtext, fatemi2024talk, guo2023gpt4graph}. These techniques show that it is possible to encode nodes, edges, and paths into token sequences in a way that preserves useful structural information and allows the model to perform tasks such as classification, question answering, or link prediction over graphs.

Our work builds on these ideas in a domain-specific way. Rather than targeting generic graphs, \textsc{Helios} focuses on the control flow and call graphs that arise in binary analysis, together with the \texttt{P-Code} level operations attached to each basic block. The framework constructs a hierarchical textual representation with three main components: a function-level summary that captures inter-function structure and coarse semantic patterns, a logical flow section that presents intra-function control flow and explicit successor relationships, and a block-level view that provides the low-level evidence. On top of this structure, \textsc{Helios} adds a small set of natural language rules that constrain how the model should use the graph information, and an optional compiler feedback loop.

In summary, prior work has established graph-based representations as central for binary analysis and has applied LLMs to decompilation and graph reasoning, but these lines of work are largely separate. Fine-tuned decompilers optimize for direct translation at the cost of architectural specialization, and prompt-based refinement tools mainly target readability while remaining structurally blind. \textsc{Helios} combines program analysis, graph-aware prompting, and compiler feedback to encourage a general-purpose LLM to reason about control flow and correctness rather than only about textual similarity, and we evaluate this design across multiple architectures without any task-specific training.

\section{The \textsc{Helios} Framework}
\label{sec:architecture}

To address the challenge of ``structural blindness" in current LLM-based decompilation, we designed and implemented \textsc{Helios}, a modular, end-to-end framework for augmenting LLMs with the rich, structural context of a binary. Instead of treating decompilation as a text-to-text translation task, \textsc{Helios} reframes it as a context-aware reasoning problem. The core principle of our framework is to provide the LLM not only with the decompiled pseudo-code but also with a textual representation that models the program's underlying graph structures at multiple levels of abstraction. This approach is designed to mimic the analytical process of a human expert, who synthesizes information from various representations to form a holistic understanding of the program.

\begin{figure}[t]
    \centering
    \includegraphics[width=0.48\textwidth]{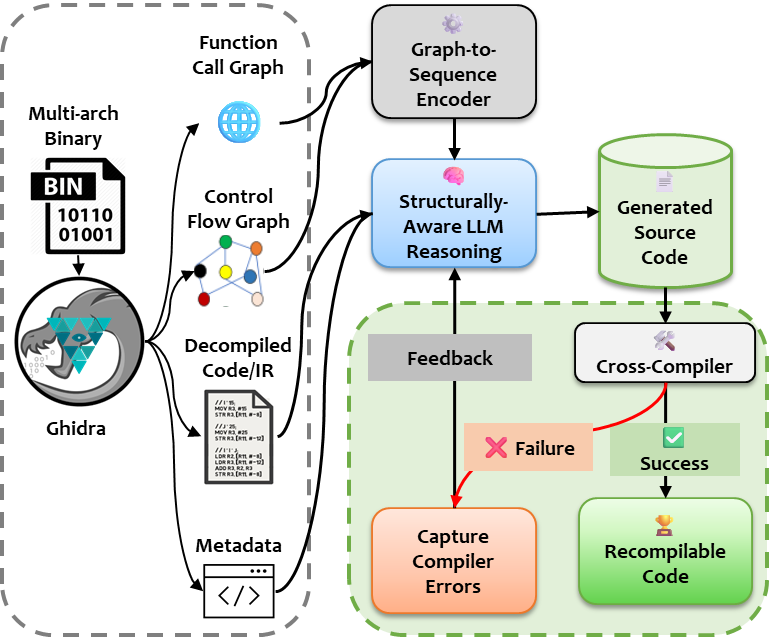} % Updated width for better fit
    % \vspace{-0.3cm}
   \caption{The high-level architecture of the \textsc{Helios} framework. A binary is first analyzed to extract the decompiled code and its CFG. \textsc{Helios} then abstracts the CFG and combines it with other metadata to form a context-rich prompt for the LLM, enabling structurally-aware decompilation and optional iterative refinement via compiler feedback.}
    \label{fig:architecture}
\end{figure}
\subsection{System Overview}

The \textsc{Helios} pipeline, shown in Figure~\ref{fig:architecture}, operates per function and follows a simple pattern. Given a binary, we run a static analysis backend to recover pseudo-C code, control-flow graphs, and related metadata. For each function, we then build a hierarchical textual summary of this structure, attach the raw decompiler output, and pass the result to a general-purpose LLM. An optional compiler-in-the-loop checks the generated code and, if necessary, triggers a single round of repairs.

For clarity, we refer to the main stages as:
\begin{enumerate}
    \item \textbf{Static analysis and feature extraction}, which collects per-function artifacts such as pseudo-code, CFG, and call information.
    \item \textbf{Hierarchical prompt generation}, which encodes these artifacts into the \textsc{Helios} prompt format.
    \item \textbf{Structurally aware prompting}, which uses a fixed instruction template plus the prompt to guide the LLM.
    \item \textbf{Iterative refinement with compiler feedback}, which is an optional second pass based on compiler diagnostics.
\end{enumerate}
The following subsections describe the analysis, prompt format, and prompting strategy in more detail.
\subsection{Static Analysis and Feature Extraction}

The first stage relies on the rich artifacts produced by modern reverse engineering platforms. Our implementation uses the Ghidra Software Reverse Engineering Framework~\cite{ghidra} and its headless analyzer, driven by a script that processes whole binaries in a reproducible way. The design is tool-agnostic and could be implemented on top of other analysis frameworks with similar capabilities.

For each function, we extract:
\begin{itemize}
    \item the decompiler's C or C-like pseudo code, including the function signature and local variables,
    \item the complete control flow graph (CFG) derived from Ghidra's \texttt{P-Code} intermediate representation~\cite{ghidra_pcode_docs}, including a list of basic blocks and directed edges between them~\cite{brumley2013native},
    \item the interprocedural function call graph (FCG), restricted to calls that originate in the current function, and
    \item auxiliary metadata such as loop headers, string references, imported functions, and constant values.
\end{itemize}

During this pass, we also record a mapping between each basic block and the corresponding region of the decompiled pseudo-code. This mapping allows the prompt generator to refer to blocks in a stable way and to cross-check CFG edges against the textual output. All subsequent stages operate on these per-function summaries rather than on raw binaries.

\subsection{Hierarchical Abstraction and Prompt Format}

\begin{figure}[t]
    \centering
    \includegraphics[width=0.48\textwidth]{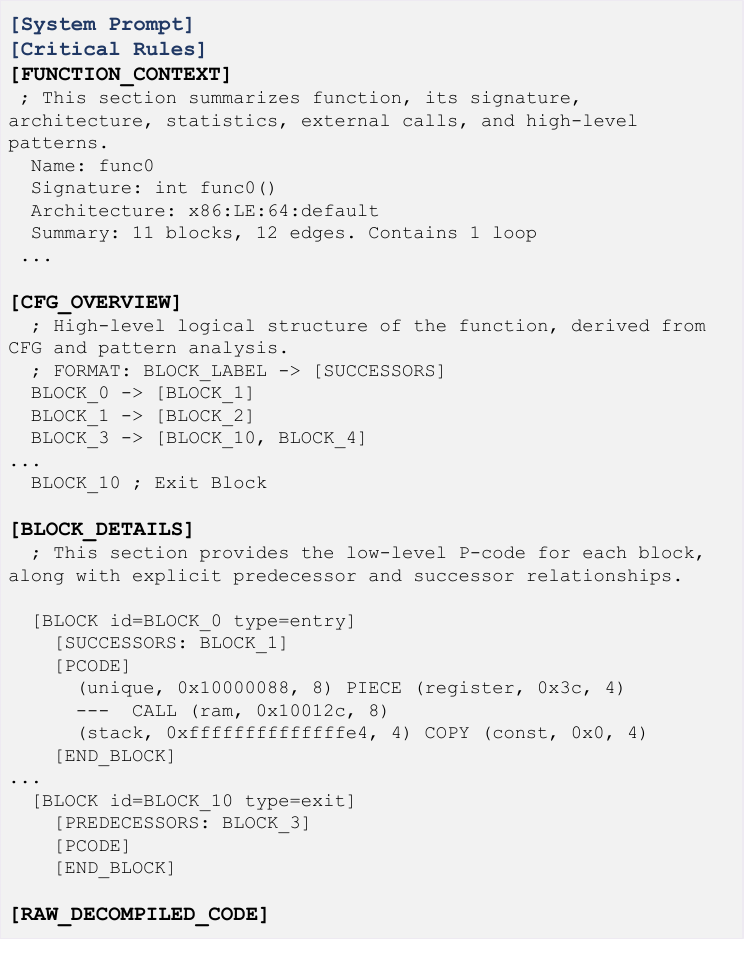}
    % \vspace{-0.3cm}
   \caption{The Multi-Part Prompt Format of \textsc{Helios}. Our prompt provides a hierarchical view of the binary, starting with a high-level \textbf{\texttt{[FUNCTION\_CONTEXT]}}, followed by the \textbf{\texttt{[CFG\_OVERVIEW]}} (a topological map of the CFG), and finally, the low-level evidence in \textbf{\texttt{[BLOCK\_DETAILS]}} before the \texttt{\textbf{[RAW\_DECOMPILED\_CODE]}}.}
\label{fig:prompt_example}
\end{figure}

The second stage converts the analysis artifacts into a structured prompt. The goal is not only to linearize the CFG, but to present the program at several levels of abstraction that mirror how a human analyst would approach it. We adopt a simple segment based format, illustrated in Figure~\ref{fig:prompt_example}, with four main sections:

\begin{enumerate}
    \item \textbf{\texttt{[FUNCTION\_CONTEXT]}} A top-level summary that includes the function name and signature, the target architecture, and coarse structural statistics such as the number of basic blocks and loops. This gives the model a quick mental picture of the function before it sees any code.
    \item \textbf{\texttt{[CFG\_OVERVIEW]}} A compact description of the CFG that lists, for each basic block, its successors and any notable role such as loop header or branch target. This section provides a readable map of the possible execution paths without exposing low level instructions.
    \item \textbf{\texttt{[BLOCK\_DETAILS]}} A block by block summary that contains distilled \texttt{P-Code} instructions for each basic block, with stable block identifiers that match those used in the cfg overview. This serves as the main source of semantic evidence for the model.
    \item \textbf{\texttt{[RAW\_DECOMPILED\_CODE]}} The original pseudo C output from the decompiler, presented without modification. This gives the model a baseline implementation to refine.
\end{enumerate}

In a typical case, \texttt{[FUNCTION\_CONTEXT]} is only a few lines, \texttt{[CFG\_OVERVIEW]} is a list of tens of basic blocks, and \texttt{[BLOCK\_DETAILS]} contributes a few hundred tokens of \texttt{P-Code}. The complete prompt for a single function remains well within the context limits of current models.

This hierarchical layout is intended to address structural blindness directly. The \texttt{[FUNCTION\_CONTEXT]} section supplies the high-level structure, such as the presence of loops or early returns. The \texttt{[CFG\_OVERVIEW]} section materializes the CFG as a simple adjacency list that the model can reason over. The \texttt{[BLOCK\_DETAILS]} section grounds that reasoning in concrete low-level operations. Finally, \texttt{[RAW\_DECOMPILED\_CODE]} anchors the output to the existing decompiler, so the model can focus on correcting and simplifying rather than inventing a completely new implementation.

\subsection{Structurally Aware Prompting and Iterative Refinement}

In the final stage, \textsc{Helios} turns the structural summary into a concrete instruction for the LLM. The prompt follows a universal template that has three parts: a short description of the task, a list of critical rules that the model should follow, and the four structural sections described above.

The critical rules are based on a manual study of common failure modes we observed when prompting LLMs with decompiler output. We grouped the failures into a small number of recurring patterns, including:
\begin{itemize}
    \item reimplementing standard library functions instead of calling them,
    \item producing control flow that does not match the CFG, for example, by inventing extra branches or omitting error paths, and
    \item introducing type mismatches, especially in code compiled with high optimization levels, where type information is partially erased.
\end{itemize}
Each pattern is expressed as a short natural language rule in the system prompt, for example, that all branches in the generated code must correspond to branches in \texttt{[CFG\_OVERVIEW]}, and that new global variables must not be introduced unless they are present in \texttt{[BLOCK\_DETAILS]} or \texttt{[RAW\_DECOMPILED\_CODE]}. The ablation study in Table~\ref{tab:rq4_ablation} shows that adding these rules on top of the structural sections yields a large jump in compilability, which suggests that guidance on how to use the structure is as important as the structure itself.

To further improve robustness, \textsc{Helios} can optionally run a compiler feedback loop. After the first model call, the generated C code is compiled with the same toolchain used to create the original binary. If compilation succeeds and the resulting object file links into a test harness, the process stops. If compilation fails, the compiler diagnostics are appended to the prompt in a dedicated section, and the model is asked to correct the code while maintaining control-flow consistency with \texttt{[CFG\_OVERVIEW]} and \texttt{[BLOCK\_DETAILS]}. This produces a second candidate, which we compile again and then evaluate with unit tests. In all reported experiments, we use at most one such feedback iteration per function, which keeps the overall cost predictable while still yielding large gains in compilability and functional correctness.

\section{Experimental Setup}
\label{sec:experiments}

To evaluate the efficacy of \textsc{Helios}, we designed a series of experiments to answer the following research questions, framed to assess both the practical security application and the underlying AI advancements:

\textbf{RQ1:} Does providing explicit structural graph context to an LLM lead to a significant improvement in the quality and correctness of decompilation compared to approaches that reason over unstructured text alone?

\textbf{RQ2:} How effectively does HELIOS generalize to different hardware architectures compared to structurally-blind methods?

\textbf{RQ3:} To what extent does an optional, iterative compiler feedback loop improve the rate of syntactically correct and re-executable code generation?

\textbf{RQ4:}  What is the individual contribution of each component in the HELIOS hierarchical prompt to the overall improvement in decompilation quality?

\subsection{Datasets}

Our experiments use two widely adopted code generation benchmarks,
\textbf{HumanEval}~\cite{chen2021evaluating} and
\textbf{MBPP}~\cite{austin2021programsynthesislargelanguage}.
Both provide natural language specifications and high-quality unit test
suites, which we use as an objective measure of functional correctness.
Following prior work on LLM-based decompilation, such as
LLM4Decompile~\cite{tan2024llm4decompile}, we use the publicly available
\textit{C-converted} versions of these benchmarks, which supply C implementations
aligned with the original Python tasks and compatible test suites.
% In this section we refer to these as the C versions of HumanEval and MBPP;we reserve the name \textbf{HumanEval-Decompile} for the dataset ofdecompiled outputs that we introduce later in the paper.

From these C programs, we construct our primary dataset, the
\textbf{Cross-Architecture Decompilation Database (Cross-Arch-DB)}.
For each task, we compile a self-contained C implementation across
\textbf{six} target architectures:
\texttt{x86\_32} (i686), \texttt{x86\_64}, \texttt{arm\_32},
\texttt{aarch64}, \texttt{mips\_32}, and \texttt{mips\_64}. These targets
cover the three most common instruction set families and span both CISC
(x86) and RISC (ARM, MIPS) designs. Each binary is compiled with GCC 11.4
using four optimization levels (\texttt{-O0} through \texttt{-O3}). The
resulting binaries and their associated test suites form \textbf{Cross-Arch-DB},
which we use to study both architecture-specific behavior (RQ1) and
cross-architecture generalization (RQ2).
When discussing results on decompiled outputs, we
refer to the HumanEval-derived test cases as \textbf{HumanEval-Decompile}
and the MBPP-derived test cases as \textbf{MBPP-Decompile}.

\subsection{Evaluation Metrics}
To evaluate the quality of the generated code, we adopt four metrics established in prior work on decompilation correctness and code generation~\cite{liu2020how, chen2021evaluating, tan2024llm4decompile, slade2024}. 
% Each metric is calculated as a percentage or score over the total number of functions in the evaluation set.

\begin{itemize}
    \item \textbf{Object File Compilability (Re-compilability):} The percentage of all functions that successfully compile into an object file (\texttt{.o}). This serves as our check for syntactic \& type correctness.

    \item \textbf{Executable Linkability:} The percentage of all functions that can be successfully linked into a complete executable binary, a stricter test that also verifies the resolution of all external symbols.

    \item \textbf{Functional Correctness (Re-executability):} The percentage of all functions that produce a valid executable \textit{and} pass their original ground-truth test suite. This is our strictest measure of semantic preservation.

    \item \textbf{Edit Similarity:} To measure the textual closeness to the original source, we use a metric based on the normalized Levenshtein edit distance. A higher score indicates a more textually accurate reconstruction~\cite{tan2024llm4decompile}.
\end{itemize}

\subsection{Baselines}
To rigorously evaluate the contributions of \textsc{Helios}, we compare it against two representative baseline categories: (1) an \textbf{LLM-Text-Only} baseline that provides the LLM with only the raw Ghidra decompiled text, and (2) \textbf{LLM-Finetuned} models, using the publicly available \textit{Nova}~\cite{jiang2023nova} and \textit{LLM4Decompile}~\cite{tan2024llm4decompile} models (6.7B and 1.3B variants) as representatives of the state-of-the-art fine-tuning approach. In line with~\cite{tan2024llm4decompile}, we do not compare against SLaDe~\cite{slade2024}, as its methodology requires intermediate compiler artifacts that are unavailable in the realistic black-box scenario our work addresses.

\subsection{Implementation Details}
Our \textsc{Helios} prototype is built using Ghidra~11.0 for static analysis and Python~3.10 for the prompt generation modules. For our experiments, we use two general-purpose LLMs, \texttt{gpt-4.1-mini} and \texttt{gemini-2.0-flash}, chosen for their balance of reasoning capability, cost, and inference speed. Experiments were orchestrated from a server with an NVIDIA A100 GPU, while all model inferences are issued via hosted APIs. We will release our implementation, prompts, and evaluation scripts, together with instructions for reconstructing Cross-Arch-DB, upon acceptance.

We report our primary experimental results, including performance across architectures, optimization levels, and evaluation metrics, in the main body of the paper. The Appendix provides additional evaluations and details, including the standardized NOVA evaluation protocol, and ablation tables.

\begin{table*}[t!]
\centering

{\small

\caption{Performance comparison on \textbf{x86\_64 architecture} across optimization levels O0, O1, and O3 on the HumanEval-Decompile dataset. Our \textsc{Helios} framework is benchmarked against baselines across four key metrics.}
\label{tab:x86_result}
}
\setlength{\tabcolsep}{5pt} 
\begin{tabular}{@{}l|cccc|cccc|cccc|cccc@{}}
\toprule
 & \multicolumn{4}{c|}{\textbf{Obj. Compilability (\%)}} & \multicolumn{4}{c|}{\textbf{Exec. Linkability (\%)}} & \multicolumn{4}{c|}{\textbf{Func. Correctness (\%)}} & \multicolumn{4}{c}{\textbf{Edit Similarity}} \\
\cmidrule(lr){2-5} \cmidrule(lr){6-9} \cmidrule(lr){10-13} \cmidrule(lr){14-17}
\textbf{Model} & \textbf{O0} & \textbf{O1} & \textbf{O3} & \textbf{AVG} & \textbf{O0} & \textbf{O1} & \textbf{O3} & \textbf{AVG} & \textbf{O0} & \textbf{O1} & \textbf{O3} & \textbf{AVG} & \textbf{O0} & \textbf{O1} & \textbf{O3} & \textbf{AVG} \\
\midrule
\multicolumn{17}{@{}l}{\textbf{A. Text-Only Baselines (Structurally-Blind)}} \\
\hspace{3mm}Gemini-2.0-Flash & 
  \cellcolor[HTML]{FEFFFE}68.6 &
  \cellcolor[HTML]{EEA7A1}40.1 &
  \cellcolor[HTML]{E67C73}26.2 &
  \cellcolor[HTML]{F2BCB8}45.0 &
  \cellcolor[HTML]{E9F6F0}60.1 &
  \cellcolor[HTML]{F7D8D6}41.4 &
  \cellcolor[HTML]{F1BAB5}30.7 &
  \cellcolor[HTML]{F5CDCA}44.1 &
  \cellcolor[HTML]{C9EADA}53.0 &
  \cellcolor[HTML]{FAE6E5}35.2 &
  \cellcolor[HTML]{F4CAC6}26.2 &
  \cellcolor[HTML]{F7D9D7}38.1 &
  \cellcolor[HTML]{F0F9F5}30.5 &
  \cellcolor[HTML]{F4C5C1}23.8 &
  \cellcolor[HTML]{EB9992}20.3 &
  \cellcolor[HTML]{F1B9B5}24.9 \\
\hspace{3mm}GPT-4.1 Mini & 
  \cellcolor[HTML]{A3DABF}85.8 &
  \cellcolor[HTML]{F1FAF5}71.0 &
  \cellcolor[HTML]{F8DCDA}57.3 &
  \cellcolor[HTML]{DAEEE3}71.4 &
  \cellcolor[HTML]{83CDA9}84.8 &
  \cellcolor[HTML]{D7EFE3}64.5 &
  \cellcolor[HTML]{FDF7F7}52.1 &
  \cellcolor[HTML]{C7E9D8}67.1 &
  \cellcolor[HTML]{57BB8A}74.3 &
  \cellcolor[HTML]{CCEBDC}52.4 &
  \cellcolor[HTML]{E8F6EF}47.2 &
  \cellcolor[HTML]{A9DDC4}58.0 &
  \cellcolor[HTML]{E8F6EF}31.6 &
  \cellcolor[HTML]{F5CAC7}24.1 &
  \cellcolor[HTML]{E67C73}18.0 &
  \cellcolor[HTML]{F0B5B0}24.6 \\
\midrule
\multicolumn{17}{@{}l}{\textbf{B. Specialized Baselines (Fine-Tuned)}} \\
\hspace{3mm}LLM4Decompile (1.3B) & 
  \cellcolor[HTML]{F2BEBA}47.6 &
  \cellcolor[HTML]{F3C2BE}48.8 &
  \cellcolor[HTML]{F1BAB6}46.3 &
  \cellcolor[HTML]{F2BEB9}47.6 &
  \cellcolor[HTML]{FBEAE9}47.6 &
  \cellcolor[HTML]{FAE7E6}46.7 &
  \cellcolor[HTML]{F8DEDC}43.3 &
  \cellcolor[HTML]{F9E3E1}45.9 &
  \cellcolor[HTML]{F9E4E2}34.5 &
  \cellcolor[HTML]{F4C8C4}25.6 &
  \cellcolor[HTML]{F0B5B0}19.5 &
  \cellcolor[HTML]{F4C6C2}26.5 &
  \cellcolor[HTML]{88CFAC}46.1 &
  \cellcolor[HTML]{B2E0C9}39.8 &
  \cellcolor[HTML]{BBE4D0}38.4 &
  \cellcolor[HTML]{A6DBC1}41.4 \\
\hspace{3mm}LLM4Decompile (6.7B) & 
  \cellcolor[HTML]{E5F5ED}73.2 &
  \cellcolor[HTML]{F9E4E2}59.8 &
  \cellcolor[HTML]{F8DBD8}56.7 &
  \cellcolor[HTML]{F4C8C4}63.2 &
  \cellcolor[HTML]{C3E7D6}69.2 &
  \cellcolor[HTML]{FEFFFE}55.2 &
  \cellcolor[HTML]{FCF3F2}50.6 &
  \cellcolor[HTML]{EBF7F1}58.3 &
  \cellcolor[HTML]{C4E8D6}54.0 &
  \cellcolor[HTML]{F7D7D4}30.2 &
  \cellcolor[HTML]{F4C5C1}24.7 &
  \cellcolor[HTML]{F5CBC7}36.3 &
  \cellcolor[HTML]{57BB8A}53.2 &
  \cellcolor[HTML]{97D5B7}43.7 &
  \cellcolor[HTML]{ADDEC6}40.5 &
  \cellcolor[HTML]{8BD0AE}45.8 \\
\hspace{3mm}Nova (1.3B) & 
  \cellcolor[HTML]{E88A82}30.8 &
  \cellcolor[HTML]{EDA39D}39.0 &
  \cellcolor[HTML]{EDA29C}38.7 &
  \cellcolor[HTML]{EB9B94}36.2 &
  \cellcolor[HTML]{E67C73}8.8 &
  \cellcolor[HTML]{EC9D96}20.4 &
  \cellcolor[HTML]{EA918A}16.5 &
  \cellcolor[HTML]{E98C84}15.2 &
  \cellcolor[HTML]{E67C73}1.2 &
  \cellcolor[HTML]{E78179}3.1 &
  \cellcolor[HTML]{E67F77}2.4 &
  \cellcolor[HTML]{E67E76}2.2 &
  \cellcolor[HTML]{F0B4AF}22.4 &
  \cellcolor[HTML]{F7D9D7}25.3 &
  \cellcolor[HTML]{F5CFCB}24.5 &
  \cellcolor[HTML]{F5CBC7}24.1 \\
\hspace{3mm}Nova (6.7B) & 
  \cellcolor[HTML]{FDF8F8}66.2 &
  \cellcolor[HTML]{F9FDFB}69.5 &
  \cellcolor[HTML]{FDF7F6}65.9 &
  \cellcolor[HTML]{FEFBFA}67.2 &
  \cellcolor[HTML]{F7D9D6}41.5 &
  \cellcolor[HTML]{F5CBC7}36.6 &
  \cellcolor[HTML]{F3C5C1}34.5 &
  \cellcolor[HTML]{F4C8C5}37.5 &
  \cellcolor[HTML]{E7847C}4.0 &
  \cellcolor[HTML]{E78179}3.1 &
  \cellcolor[HTML]{E67F77}2.4 &
  \cellcolor[HTML]{E78178}3.2 &
  \cellcolor[HTML]{EBF7F1}31.2 &
  \cellcolor[HTML]{EAF7F0}31.4 &
  \cellcolor[HTML]{F8FCFA}29.4 &
  \cellcolor[HTML]{F1F9F5}30.7  \\
\midrule
\multicolumn{17}{@{}l}{\textbf{C. Our Method (\textsc{Helios} w/ Structural Context)}} \\
\hspace{3mm}Gemini-2.0-Flash + \textsc{Helios} & 
  \cellcolor[HTML]{64C193}97.6 &
  \cellcolor[HTML]{C2E7D5}79.8 &
  \cellcolor[HTML]{CAEADA}78.3 &
  \cellcolor[HTML]{94D4B5}85.2 &
  \cellcolor[HTML]{BBE4D0}71.3 &
  \cellcolor[HTML]{FEFEFE}54.4 &
  \cellcolor[HTML]{FDF9F9}52.8 &
  \cellcolor[HTML]{EDCDC9}59.5 &
  \cellcolor[HTML]{AEDFC7}58.1 &
  \cellcolor[HTML]{EAF7F1}46.9 &
  \cellcolor[HTML]{FEFEFE}42.7 &
  \cellcolor[HTML]{E6F5EE}49.2 &
  \cellcolor[HTML]{CBEADB}36.0 &
  \cellcolor[HTML]{F3FAF7}30.1 &
  \cellcolor[HTML]{F2BCB7}23.0 &
  \cellcolor[HTML]{F8DDDB}29.7 \\
\hspace{3mm}GPT-4.1 Mini + \textsc{Helios} & 
  \cellcolor[HTML]{6AC397}96.6 &
  \cellcolor[HTML]{AFDFC7}83.6 &
  \cellcolor[HTML]{94D4B5}88.6 &
  \cellcolor[HTML]{86CFAB}89.6 &
  \cellcolor[HTML]{A6DBC1}76.4 &
  \cellcolor[HTML]{D2EDE0}65.8 &
  \cellcolor[HTML]{EBF7F1}59.6 &
  \cellcolor[HTML]{C7E9D8}67.3 &
  \cellcolor[HTML]{8ED1B0}64.2 &
  \cellcolor[HTML]{F1FAF5}45.6 &
  \cellcolor[HTML]{FDF9F9}41.1 &
  \cellcolor[HTML]{CBE7DA}50.3 &
  \cellcolor[HTML]{C6E8D7}36.8 &
  \cellcolor[HTML]{FEFAFA}27.8 &
  \cellcolor[HTML]{F2BFBB}23.3 &
  \cellcolor[HTML]{F7D8D6}29.3 \\
\hspace{3mm}Gemini-2.0-Flash + w/ Fback & 
  \cellcolor[HTML]{57BB8A}100.0 &
  \cellcolor[HTML]{7FCCA6}92.5 &
  \cellcolor[HTML]{81CCA7}92.2 &
  \cellcolor[HTML]{6EC59A}94.9 &
  \cellcolor[HTML]{84CEAA}84.5 &
  \cellcolor[HTML]{C7E9D8}68.4 &
  \cellcolor[HTML]{E1F3EA}62.1 &
  \cellcolor[HTML]{B4E1CA}71.7 &
  \cellcolor[HTML]{81CCA7}66.6 &
  \cellcolor[HTML]{E5F5ED}47.9 &
  \cellcolor[HTML]{F4FBF8}45.0 &
  \cellcolor[HTML]{D5EEE2}53.2 &
  \cellcolor[HTML]{E8F6EF}31.6 &
  \cellcolor[HTML]{F5CAC7}24.1 &
  \cellcolor[HTML]{E67C73}18.0 &
  \cellcolor[HTML]{F1B8B4}24.6  \\
\hspace{3mm}GPT-4.1 Mini + w/ Feedback & 
  \cellcolor[HTML]{5BBD8D}99.3 &
  \cellcolor[HTML]{7CCAA4}93.1 &
  \cellcolor[HTML]{67C295}97.1 &
  \cellcolor[HTML]{6AC397}96.5 &
  \cellcolor[HTML]{57BB8A}95.3 &
  \cellcolor[HTML]{8DD1B0}82.4 &
  \cellcolor[HTML]{C3E7D6}69.3 &
  \cellcolor[HTML]{85CEAA}82.3 &
  \cellcolor[HTML]{71C69C}69.6 &
  \cellcolor[HTML]{CEECDD}52.1 &
  \cellcolor[HTML]{EFF9F4}46.0 &
  \cellcolor[HTML]{ABDDC5}55.9 &
  \cellcolor[HTML]{C6E8D7}36.8 &
  \cellcolor[HTML]{FEFAFA}27.8 &
  \cellcolor[HTML]{F2BFBB}23.3 &
  \cellcolor[HTML]{F7D8D6}29.3\\
\bottomrule
\end{tabular}
\end{table*}

\section{Results and Analysis}
\label{sec:results}

In this section, we present the results of our experiments, organized by research question. Overall, our findings show that providing hierarchical structural context allows general-purpose LLMs to produce substantially more reliable decompilations than text-only prompts or specialized fine-tuned models.

\begin{table*}[t!]
\centering
\caption{
Detailed performance on the \texttt{MBPP} dataset for the x86\_64 architecture across compiler optimization levels (\texttt{-O0}, \texttt{-O1}, \texttt{-O3}). \textsc{Helios} consistently outperforms structurally-blind baselines and LLM4Decompile variants across all metrics. Feedback-based refinement notably boosts cross-stage consistency.
}
{\small
\setlength{\tabcolsep}{2.5pt} % Slightly more compact
\begin{tabular}{@{}l|cccc|cccc|cccc|cccc@{}}
\toprule
& \multicolumn{4}{c|}{\textbf{Obj. Compilability (\%)}} & \multicolumn{4}{c|}{\textbf{Exec. Linkability (\%)}} & \multicolumn{4}{c|}{\textbf{Func. Correctness (\%)}} & \multicolumn{4}{c}{\textbf{Edit Similarity}} \\
\cmidrule(lr){2-5} \cmidrule(lr){6-9} \cmidrule(lr){10-13} \cmidrule(lr){14-17}
\textbf{Model} & \textbf{O0} & \textbf{O1} & \textbf{O3} & \textbf{AVG} & \textbf{O0} & \textbf{O1} & \textbf{O3} & \textbf{AVG} & \textbf{O0} & \textbf{O1} & \textbf{O3} & \textbf{AVG} & \textbf{O0} & \textbf{O1} & \textbf{O3} & \textbf{AVG} \\
\midrule
\multicolumn{17}{@{}l}{\textit{Gemini-2.0-Flash}} \\
\hspace{3mm}Baseline &
\cellcolor[HTML]{FEFFFE}62.65 & \cellcolor[HTML]{E98E87}46.28 & \cellcolor[HTML]{E67C73}35.46 & \cellcolor[HTML]{EB9A93}45.87 &
\cellcolor[HTML]{FFFFFF}62.65 & \cellcolor[HTML]{FFFFFF}46.28 & \cellcolor[HTML]{FFFFFF}35.46 & \cellcolor[HTML]{FFFFFF}45.87 &
\cellcolor[HTML]{FFFFFF}57.16 & \cellcolor[HTML]{FFFFFF}41.11 & \cellcolor[HTML]{FFFFFF}31.65 & \cellcolor[HTML]{FFFFFF}41.36 &
\cellcolor[HTML]{E67C73}34.29 & \cellcolor[HTML]{EEA7A1}28.19 & \cellcolor[HTML]{FFFFFF}21.84 & \cellcolor[HTML]{EB9992}27.46 \\

\hspace{3mm}HELIOS &
\cellcolor[HTML]{AADDC4}81.25 & \cellcolor[HTML]{D2EDE0}62.76 & \cellcolor[HTML]{DAF0E5}57.14 & \cellcolor[HTML]{C9E9DA}64.90 &
\cellcolor[HTML]{71C69C}81.25 & \cellcolor[HTML]{97D5B6}62.76 & \cellcolor[HTML]{7CCAA4}57.14 & \cellcolor[HTML]{82CDA8}64.89 &
\cellcolor[HTML]{E3F4EC}59.73 & \cellcolor[HTML]{DBF1E6}44.97 & \cellcolor[HTML]{C0E6D3}38.39 & \cellcolor[HTML]{CEECDD}46.35 &
\cellcolor[HTML]{FAE9E7}37.49 & \cellcolor[HTML]{E67C73}27.35 & \cellcolor[HTML]{E67C73}19.86 & \cellcolor[HTML]{E67C73}27.14 \\

\hspace{3mm}HELIOS w/ Feedback &
\cellcolor[HTML]{57BB8A}99.00 & \cellcolor[HTML]{57BB8A}95.48 & \cellcolor[HTML]{57BB8A}93.99 & \cellcolor[HTML]{57BB8A}95.96 &
\cellcolor[HTML]{57BB8A}84.53 & \cellcolor[HTML]{57BB8A}72.67 & \cellcolor[HTML]{57BB8A}63.08 & \cellcolor[HTML]{57BB8A}71.39 &
\cellcolor[HTML]{57BB8A}72.42 & \cellcolor[HTML]{57BB8A}58.82 & \cellcolor[HTML]{57BB8A}49.37 & \cellcolor[HTML]{57BB8A}58.42 &
\cellcolor[HTML]{FFFFFF}38.12 & \cellcolor[HTML]{FFFFFF}29.86 & \cellcolor[HTML]{F2BEB9}20.86 & \cellcolor[HTML]{FFFFFF}28.56 \\
\midrule
\hspace{3mm}LLM4Decompile (1.3B) &
\cellcolor[HTML]{E67C73}46.46 & \cellcolor[HTML]{E67C73}45.53 & \cellcolor[HTML]{F6CFCC}42.51 & \cellcolor[HTML]{E67C73}44.60 &
\cellcolor[HTML]{E67C73}32.19 & \cellcolor[HTML]{E67C73}26.13 & \cellcolor[HTML]{E67C73}22.48 & \cellcolor[HTML]{E67C73}26.18 &
\cellcolor[HTML]{E67C73}32.19 & \cellcolor[HTML]{E67C73}26.13 & \cellcolor[HTML]{E67C73}22.48 & \cellcolor[HTML]{E67C73}26.18 &
\cellcolor[HTML]{96D5B6}47.11 & \cellcolor[HTML]{65C194}41.81 & \cellcolor[HTML]{5DBE8E}36.04 & \cellcolor[HTML]{6EC59A}41.11 \\

\hspace{3mm}LLM4Decompile (6.7B) &
\cellcolor[HTML]{F1B6B1}53.70 & \cellcolor[HTML]{FFFFFF}50.77 & \cellcolor[HTML]{FFFFFF}46.46 & \cellcolor[HTML]{FFFFFF}50.02 &
\cellcolor[HTML]{EEA9A3}42.71 & \cellcolor[HTML]{F0B2AD}34.55 & \cellcolor[HTML]{F2BEBA}29.11 & \cellcolor[HTML]{F0B3AE}34.52 &
\cellcolor[HTML]{F0B3AD}42.71 & \cellcolor[HTML]{F4C5C1}34.55 & \cellcolor[HTML]{F8DAD8}29.11 & \cellcolor[HTML]{F3C3BF}34.52 &
\cellcolor[HTML]{57BB8A}52.40 & \cellcolor[HTML]{57BB8A}42.88 & \cellcolor[HTML]{57BB8A}36.49 & \cellcolor[HTML]{57BB8A}43.03 \\
\bottomrule
\end{tabular}

\label{tab:mbpp_x86_results}
}

\end{table*}

\begin{table*}[t!]
\centering
\scriptsize
\caption{Comprehensive performance results across multiple architectures for Obj. Compilability (\%), Exec. Linkability (\%), Func. Correctness (\%), and Edit Similarity across all optimization levels (-O0 to -O3) on HumanEval-Decompile.}
\setlength{\tabcolsep}{2.2pt}
{%
\begin{tabular}{@{}l|ccccc|ccccc|ccccc|ccccc@{}}
\toprule
 & \multicolumn{5}{c|}{\textbf{Obj. Compilability (\%)}} & \multicolumn{5}{c|}{\textbf{Exec. Linkability (\%)}} & \multicolumn{5}{c|}{\textbf{Func. Correctness (\%)}} & \multicolumn{5}{c}{\textbf{Edit Similarity}} \\
\cmidrule(lr){2-6} \cmidrule(lr){7-11} \cmidrule(lr){12-16} \cmidrule(lr){17-21}
\textbf{Model} & \textbf{O0} & \textbf{O1} & \textbf{O2} & \textbf{O3} & \textbf{AVG} & \textbf{O0} & \textbf{O1} & \textbf{O2} & \textbf{O3} & \textbf{AVG} & \textbf{O0} & \textbf{O1} & \textbf{O2} & \textbf{O3} & \textbf{AVG} & \textbf{O0} & \textbf{O1} & \textbf{O2} & \textbf{O3} & \textbf{AVG} \\
\midrule
\cmidrule(lr){2-6} \cmidrule(lr){7-11} \cmidrule(lr){12-16} \cmidrule(lr){17-21}
\midrule
\multicolumn{21}{@{}c}{\textbf{ARM32}} \\
\midrule
\multicolumn{21}{@{}l}{\textit{Gemini-2.0-Flash}} \\
\hspace{3mm}Baseline & \cellcolor[HTML]{E67C73}60.88 & \cellcolor[HTML]{E67C73}46.71 & \cellcolor[HTML]{E67C73}37.25 & \cellcolor[HTML]{E67C73}37.91 & \cellcolor[HTML]{E67C73}45.69 & \cellcolor[HTML]{E67C73}61.22 & \cellcolor[HTML]{E67C73}43.09 & \cellcolor[HTML]{E67C73}37.25 & \cellcolor[HTML]{E67C73}36.93 & \cellcolor[HTML]{E67C73}44.62 & \cellcolor[HTML]{E67C73}50.00 & \cellcolor[HTML]{E67C73}35.20 & \cellcolor[HTML]{E67C73}28.43 & \cellcolor[HTML]{E67C73}24.18 & \cellcolor[HTML]{E67C73}34.45 & \cellcolor[HTML]{E67C73}28.40 & \cellcolor[HTML]{FCF4F3}23.29 & \cellcolor[HTML]{FEFEFD}21.99 & \cellcolor[HTML]{FFFFFF}19.63 & \cellcolor[HTML]{F5CDCA}23.33 \\
\hspace{3mm}\textsc{Helios} & \cellcolor[HTML]{9FD9BD}98.64 & \cellcolor[HTML]{DEF2E8}85.86 & \cellcolor[HTML]{DFF2E9}84.64 & \cellcolor[HTML]{CAEADA}83.01 & \cellcolor[HTML]{D6EFE2}88.04 & \cellcolor[HTML]{FAE5E3}76.53 & \cellcolor[HTML]{F8DBD9}55.59 & \cellcolor[HTML]{FBEEEC}54.90 & \cellcolor[HTML]{FAE6E4}51.96 & \cellcolor[HTML]{FAE7E5}59.75 & \cellcolor[HTML]{F5CAC7}58.50 & \cellcolor[HTML]{EDA49E}36.51 & \cellcolor[HTML]{FCF4F3}37.58 & \cellcolor[HTML]{B9E3CF}37.58 & \cellcolor[HTML]{F9DFDD}40.91 & \cellcolor[HTML]{F1B6B1}29.63 & \cellcolor[HTML]{E67C73}21.30 & \cellcolor[HTML]{E67C73}19.92 & \cellcolor[HTML]{E67C73}18.11 & \cellcolor[HTML]{E67C73}22.24 \\
\hspace{3mm}\textsc{Helios} w/ Feedback & \cellcolor[HTML]{57BB8A}99.66 & \cellcolor[HTML]{57BB8A}96.38 & \cellcolor[HTML]{57BB8A}93.46 & \cellcolor[HTML]{57BB8A}92.48 & \cellcolor[HTML]{57BB8A}95.50 & \cellcolor[HTML]{57BB8A}88.78 & \cellcolor[HTML]{57BB8A}69.08 & \cellcolor[HTML]{57BB8A}68.30 & \cellcolor[HTML]{57BB8A}66.01 & \cellcolor[HTML]{57BB8A}73.04 & \cellcolor[HTML]{E9F6F0}64.97 & \cellcolor[HTML]{F3C2BE}37.50 & \cellcolor[HTML]{E8F6EF}39.22 & \cellcolor[HTML]{F8DDDB}31.37 & \cellcolor[HTML]{FAFDFB}43.26 & \cellcolor[HTML]{F7D6D3}30.30 & \cellcolor[HTML]{F7FCFA}23.64 & \cellcolor[HTML]{FAE9E7}21.66 & \cellcolor[HTML]{FEFDFD}19.59 & \cellcolor[HTML]{FCF0EF}23.80 \\
\midrule
\multicolumn{21}{@{}l}{\textit{GPT-4.1 Mini}} \\
\hspace{3mm}Baseline & \cellcolor[HTML]{F3C4C0}80.95 & \cellcolor[HTML]{F5CBC7}68.75 & \cellcolor[HTML]{F6D3D1}67.65 & \cellcolor[HTML]{F7D8D5}66.67 & \cellcolor[HTML]{F5CFCB}71.01 & \cellcolor[HTML]{DEF2E8}81.97 & \cellcolor[HTML]{C1E6D4}63.49 & \cellcolor[HTML]{D7EFE3}60.13 & \cellcolor[HTML]{C9EADA}58.82 & \cellcolor[HTML]{CCEBDC}66.10 & \cellcolor[HTML]{57BB8A}70.41 & \cellcolor[HTML]{57BB8A}52.30 & \cellcolor[HTML]{57BB8A}44.12 & \cellcolor[HTML]{57BB8A}42.81 & \cellcolor[HTML]{57BB8A}52.41 & \cellcolor[HTML]{5DBE8F}33.05 & \cellcolor[HTML]{57BB8A}26.96 & \cellcolor[HTML]{57BB8A}25.28 & \cellcolor[HTML]{57BB8A}23.19 & \cellcolor[HTML]{57BB8A}27.12 \\
\hspace{3mm}\textsc{Helios} & \cellcolor[HTML]{E7F6E2}96.94 & \cellcolor[HTML]{FDF5F4}80.59 & \cellcolor[HTML]{FDF8F8}80.39 & \cellcolor[HTML]{FCF1F0}74.51 & \cellcolor[HTML]{FDF6F6}83.11 & \cellcolor[HTML]{FCF3F2}78.57 & \cellcolor[HTML]{FAE5E4}56.91 & \cellcolor[HTML]{F9E1DF}52.94 & \cellcolor[HTML]{F9E4E2}51.63 & \cellcolor[HTML]{FAE9E7}60.01 & \cellcolor[HTML]{FDF7F6}63.27 & \cellcolor[HTML]{E6F5EE}41.45 & \cellcolor[HTML]{F6D1CE}34.97 & \cellcolor[HTML]{F6D4D1}30.72 & \cellcolor[HTML]{FEF9F9}42.60 & \cellcolor[HTML]{B7E2CD}32.00 & \cellcolor[HTML]{FDF4F4}23.30 & \cellcolor[HTML]{FFFFFF}22.02 & \cellcolor[HTML]{FBEAE9}19.38 & \cellcolor[HTML]{F5FBF8}24.18 \\
\hspace{3mm}\textsc{Helios} w/ Feedback & \cellcolor[HTML]{E7F6EF}97.62 & \cellcolor[HTML]{BCE4D1}88.49 & \cellcolor[HTML]{CBEADB}85.95 & \cellcolor[HTML]{CEEBDD}82.68 & \cellcolor[HTML]{CBEADB}88.69 & \cellcolor[HTML]{94D4B5}85.71 & \cellcolor[HTML]{8FD2B1}66.12 & \cellcolor[HTML]{BDE5D1}61.76 & \cellcolor[HTML]{BFE5D2}59.48 & \cellcolor[HTML]{A8DCC2}68.27 & \cellcolor[HTML]{85CEAA}68.71 & \cellcolor[HTML]{90D2B2}48.03 & \cellcolor[HTML]{D4EEE1}39.87 & \cellcolor[HTML]{D2EDE0}36.28 & \cellcolor[HTML]{A2DABE}48.22 & \cellcolor[HTML]{57BB8A}33.12 & \cellcolor[HTML]{BFE5D3}24.81 & \cellcolor[HTML]{C2E7D5}23.20 & \cellcolor[HTML]{E5F5ED}20.18 & \cellcolor[HTML]{B8E2CD}25.33 \\
\midrule
\multicolumn{21}{@{}c}{\textbf{mips\_32}} \\
\midrule
\multicolumn{21}{@{}l}{\textit{Gemini-2.0-Flash}} \\
\hspace{3mm}Baseline &\cellcolor[HTML]{E67C73}55.78 & \cellcolor[HTML]{E67C73}38.76 & \cellcolor[HTML]{E67C73}34.95 & \cellcolor[HTML]{E67C73}34.63 & \cellcolor[HTML]{E67C73}41.03 & \cellcolor[HTML]{E67C73}54.76 & \cellcolor[HTML]{E67C73}37.46 & \cellcolor[HTML]{E67C73}34.30 & \cellcolor[HTML]{E67C73}34.63 & \cellcolor[HTML]{E67C73}40.29 & \cellcolor[HTML]{E67C73}46.26 & \cellcolor[HTML]{E67C73}31.92 & \cellcolor[HTML]{E67C73}27.51 & \cellcolor[HTML]{E67C73}27.18 & \cellcolor[HTML]{E67C73}33.22 & \cellcolor[HTML]{E67C73}29.19 & \cellcolor[HTML]{FBEBEA}22.79 & \cellcolor[HTML]{FEFEFD}22.09 & \cellcolor[HTML]{F6D4D1}20.74 & \cellcolor[HTML]{F3C0BC}23.70
 \\
\hspace{3mm}\textsc{Helios} & \cellcolor[HTML]{7BCAA3}95.24 & \cellcolor[HTML]{D2EDE0}85.34 & \cellcolor[HTML]{D3EEE1}82.85 & \cellcolor[HTML]{CCEBDB}82.20 & \cellcolor[HTML]{CBEADB}86.41 & \cellcolor[HTML]{FDF6F6}71.43 & \cellcolor[HTML]{FBEEEC}53.75 & \cellcolor[HTML]{FCF3F2}53.40 & \cellcolor[HTML]{FAE7E5}53.72 & \cellcolor[HTML]{FCF1F1}58.08 & \cellcolor[HTML]{FBEAE8}58.84 & \cellcolor[HTML]{FCF4F3}43.00 & \cellcolor[HTML]{FEFAF9}40.13 & \cellcolor[HTML]{FDF8F8}39.81 & \cellcolor[HTML]{FCF3F2}45.44 & \cellcolor[HTML]{EFB0AA}29.90 & \cellcolor[HTML]{E67C73}22.06 & \cellcolor[HTML]{E67C73}20.47 & \cellcolor[HTML]{E67C73}19.72 & \cellcolor[HTML]{E67C73}23.04\\
\hspace{3mm}\textsc{Helios} w/ Feedback & \cellcolor[HTML]{57BB8A}99.32 & \cellcolor[HTML]{57BB8A}94.14 & \cellcolor[HTML]{57BB8A}93.53 & \cellcolor[HTML]{57BB8A}92.56 & \cellcolor[HTML]{57BB8A}94.89 & \cellcolor[HTML]{57BB8A}83.33 & \cellcolor[HTML]{57BB8A}65.80 & \cellcolor[HTML]{57BB8A}66.02 & \cellcolor[HTML]{57BB8A}66.34 & \cellcolor[HTML]{57BB8A}70.37 & \cellcolor[HTML]{9ED8BB}63.61 & \cellcolor[HTML]{D9F0E5}44.95 & \cellcolor[HTML]{F3FBF7}41.10 & \cellcolor[HTML]{C8E9D9}41.42 & \cellcolor[HTML]{CDEBDC}47.77 & \cellcolor[HTML]{FAE9E7}30.98 & \cellcolor[HTML]{EFF9F4}23.96 & \cellcolor[HTML]{FDF5F4}21.85 & \cellcolor[HTML]{D3EEE1}21.11 & \cellcolor[HTML]{F6FCF9}24.48 \\
\midrule
\multicolumn{21}{@{}l}{\textit{GPT-4.1 Mini}} \\
\hspace{3mm}Baseline & \cellcolor[HTML]{F4C6C2}78.23 & \cellcolor[HTML]{F5CBC8}65.80 & \cellcolor[HTML]{F7D6D3}66.02 & \cellcolor[HTML]{F7D8D5}63.75 & \cellcolor[HTML]{F6D0CD}68.45 & \cellcolor[HTML]{7FCBA6}75.51 & \cellcolor[HTML]{86CEAB}60.59 & \cellcolor[HTML]{C0E6D3}58.58 & \cellcolor[HTML]{CAEADA}53.72 & \cellcolor[HTML]{A8DCC2}62.10 & \cellcolor[HTML]{57BB8A}65.31 & \cellcolor[HTML]{57BB8A}48.21 & \cellcolor[HTML]{57BB8A}47.25 & \cellcolor[HTML]{DAF0E6}41.10 & \cellcolor[HTML]{57BB8A}50.47 &\cellcolor[HTML]{57BB8A}33.50 & \cellcolor[HTML]{57BB8A}26.81 & \cellcolor[HTML]{57BB8A}25.59 & \cellcolor[HTML]{57BB8A}24.50 & \cellcolor[HTML]{57BB8A}27.60
 \\
\hspace{3mm}\textsc{Helios} & \cellcolor[HTML]{D3EEE1}91.84 & \cellcolor[HTML]{FDF6F6}76.87 & \cellcolor[HTML]{FEFBFB}77.35 & \cellcolor[HTML]{FCF1F0}77.35 & \cellcolor[HTML]{FDF8F8}80.85 & \cellcolor[HTML]{E2F4EB}69.05 & \cellcolor[HTML]{FEFAF9}50.16 & \cellcolor[HTML]{FAE6E4}50.16 & \cellcolor[HTML]{FAE5E3}51.13 & \cellcolor[HTML]{FCF3F3}55.13 & \cellcolor[HTML]{F8DEDB}57.48 & \cellcolor[HTML]{F9E2E0}41.37 & \cellcolor[HTML]{F9E3E1}37.86 & \cellcolor[HTML]{FAE5E3}37.86 & \cellcolor[HTML]{F9E2E0}43.65 & \cellcolor[HTML]{C9E9D9}31.58 & \cellcolor[HTML]{FBECEB}23.23 & \cellcolor[HTML]{FFFFFF}22.23 & \cellcolor[HTML]{F4C5C1}21.45 & \cellcolor[HTML]{FDF5F4}24.62
 \\
\hspace{3mm}\textsc{Helios} w/ Feedback & \cellcolor[HTML]{E7F6E2}95.24 & \cellcolor[HTML]{DFF2E9}85.02 & \cellcolor[HTML]{F1FAF5}82.52 & \cellcolor[HTML]{BEE5D2}84.14 & \cellcolor[HTML]{E1F3EA}86.73 & \cellcolor[HTML]{FDF6F6}76.53 & \cellcolor[HTML]{F5FBF8}57.98 & \cellcolor[HTML]{E2F3EB}58.25 & \cellcolor[HTML]{CDEBDD}58.58 & \cellcolor[HTML]{E5F5ED}62.84 & \cellcolor[HTML]{81CCA8}64.29 & \cellcolor[HTML]{98D6B7}46.58 & \cellcolor[HTML]{C2E7D5}43.04 & \cellcolor[HTML]{57BB8A}43.37 & \cellcolor[HTML]{89D0AD}49.32 & \cellcolor[HTML]{79C9A2}32.74 & \cellcolor[HTML]{E0F3EA}24.25 & \cellcolor[HTML]{C1E6D4}23.23 & \cellcolor[HTML]{84CDA9}22.45 & \cellcolor[HTML]{AFDFC7}25.67 \\
\midrule
\multicolumn{21}{@{}c}{\textbf{X86\_32}} \\
\midrule
\multicolumn{21}{@{}l}{\textit{Gemini-2.0-Flash}} \\
\hspace{3mm}Baseline & \cellcolor[HTML]{E67C73}60.20 & \cellcolor[HTML]{E67C73}42.02 & \cellcolor[HTML]{E67C73}33.87 & \cellcolor[HTML]{E67C73}33.33 & \cellcolor[HTML]{E67C73}42.36 & \cellcolor[HTML]{E67C73}60.86 & \cellcolor[HTML]{E67C73}43.97 & \cellcolor[HTML]{E67C73}37.10 & \cellcolor[HTML]{E67C73}36.57 & \cellcolor[HTML]{E67C73}44.63 & \cellcolor[HTML]{E67C73}53.04 & \cellcolor[HTML]{E67C73}35.18 & \cellcolor[HTML]{E67C73}27.18 & \cellcolor[HTML]{E67C73}26.21 & \cellcolor[HTML]{E67C73}35.40 & \cellcolor[HTML]{E67C73}29.33 & \cellcolor[HTML]{F3C2BE}21.74 & \cellcolor[HTML]{FAE9E8}22.46 & \cellcolor[HTML]{FAE9E8}22.00 & \cellcolor[HTML]{F5CFCC}23.88 \\
\hspace{3mm}\textsc{Helios} & \cellcolor[HTML]{DCF1E7}94.74 & \cellcolor[HTML]{F6FCF9}80.13 & \cellcolor[HTML]{FEFAFA}81.61 & \cellcolor[HTML]{FEF9F9}81.23 & \cellcolor[HTML]{FEFDFD}84.43 & \cellcolor[HTML]{F8DBD9}76.64 & \cellcolor[HTML]{F5CECB}55.37 & \cellcolor[HTML]{F8DDDA}51.94 & \cellcolor[HTML]{FBEBEA}54.37 & \cellcolor[HTML]{F8DCDA}59.58 & \cellcolor[HTML]{F0B1AC}58.11 & \cellcolor[HTML]{FEF9F9}46.91 & \cellcolor[HTML]{FEFBFA}43.04 & \cellcolor[HTML]{FDF6F6}42.72 & \cellcolor[HTML]{FBECEB}47.69 & \cellcolor[HTML]{EC9D96}29.92 & \cellcolor[HTML]{E67C73}20.77 & \cellcolor[HTML]{E67C73}19.64 & \cellcolor[HTML]{E67C73}19.45 & \cellcolor[HTML]{E67C73}22.45 \\
\hspace{3mm}\textsc{Helios} w/ Feedback & \cellcolor[HTML]{E5F5ED}94.41 & \cellcolor[HTML]{91D3B2}85.34 & \cellcolor[HTML]{93D4B4}89.68 & \cellcolor[HTML]{83CDA9}90.61 & \cellcolor[HTML]{9DD8BB}90.01 & \cellcolor[HTML]{80CCA6}85.53 & \cellcolor[HTML]{57BB8A}66.78 & \cellcolor[HTML]{57BB8A}62.58 & \cellcolor[HTML]{57BB8A}64.08 & \cellcolor[HTML]{57BB8A}69.74 & \cellcolor[HTML]{E9F7F0}66.55 & \cellcolor[HTML]{EFF9F4}47.88 & \cellcolor[HTML]{F8FCFA}43.69 & \cellcolor[HTML]{C7E9D8}44.98 & \cellcolor[HTML]{DEF2E8}50.78 & \cellcolor[HTML]{81CCA7}33.15 & \cellcolor[HTML]{ABDDC4}24.00 & \cellcolor[HTML]{5DBE8F}24.36 & \cellcolor[HTML]{68C296}23.81 & \cellcolor[HTML]{80CCA7}26.33 \\
\midrule
\multicolumn{21}{@{}l}{\textit{GPT-4.1 Mini}} \\
\hspace{3mm}Baseline & \cellcolor[HTML]{FAE7E6}87.50 & \cellcolor[HTML]{FAE5E3}72.31 & \cellcolor[HTML]{F3C2BE}60.32 & \cellcolor[HTML]{F2BEB9}58.58 & \cellcolor[HTML]{F6D0CC}69.68 & \cellcolor[HTML]{B6E2CC}84.21 & \cellcolor[HTML]{6FC59B}66.12 & \cellcolor[HTML]{E2F4EB}58.06 & \cellcolor[HTML]{FEF9F9}56.63 & \cellcolor[HTML]{D2EDE0}66.26 & \cellcolor[HTML]{57BB8A}74.32 & \cellcolor[HTML]{57BB8A}52.44 & \cellcolor[HTML]{68C296}46.60 & \cellcolor[HTML]{57BB8A}47.25 & \cellcolor[HTML]{57BB8A}55.15 & \cellcolor[HTML]{57BB8A}33.64 & \cellcolor[HTML]{57BB8A}25.41 & \cellcolor[HTML]{57BB8A}24.41 & \cellcolor[HTML]{57BB8A}23.95 & \cellcolor[HTML]{57BB8A}26.85 \\
\hspace{3mm}\textsc{Helios} & \cellcolor[HTML]{FEFBFA}92.43 & \cellcolor[HTML]{FEFDFD}79.15 & \cellcolor[HTML]{E4F5ED}84.84 & \cellcolor[HTML]{DFF2E9}85.11 & \cellcolor[HTML]{F6FCF9}85.38 & \cellcolor[HTML]{FCF3F3}80.59 & \cellcolor[HTML]{FBEDEC}59.61 & \cellcolor[HTML]{FDF8F8}56.13 & \cellcolor[HTML]{EBF7F1}58.25 & \cellcolor[HTML]{FDF6F6}63.65 & \cellcolor[HTML]{FCF2F1}64.19 & \cellcolor[HTML]{FBEBEA}45.60 & \cellcolor[HTML]{FEFDFD}43.37 & \cellcolor[HTML]{FBEAE9}41.10 & \cellcolor[HTML]{FDF4F4}48.56 & \cellcolor[HTML]{B1E0C9}32.58 & \cellcolor[HTML]{E8F6EF}22.96 & \cellcolor[HTML]{BEE5D2}23.55 & \cellcolor[HTML]{C6E8D7}23.00 & \cellcolor[HTML]{BFE6D3}25.52 \\
\hspace{3mm}\textsc{Helios} w/ Feedback & \cellcolor[HTML]{57BB8A}99.67 & \cellcolor[HTML]{57BB8A}88.27 & \cellcolor[HTML]{57BB8A}93.23 & \cellcolor[HTML]{57BB8A}93.20 & \cellcolor[HTML]{57BB8A}93.59 & \cellcolor[HTML]{57BB8A}86.51 & \cellcolor[HTML]{A9DCC3}64.50 & \cellcolor[HTML]{A7DBC2}60.00 & \cellcolor[HTML]{B2E0C9}60.52 & \cellcolor[HTML]{99D6B8}67.88 & \cellcolor[HTML]{B0DFC8}69.60 & \cellcolor[HTML]{62C092}52.12 & \cellcolor[HTML]{57BB8A}46.93 & \cellcolor[HTML]{97D5B7}45.96 & \cellcolor[HTML]{86CEAB}53.65 & \cellcolor[HTML]{F4CAC6}30.71 & \cellcolor[HTML]{F9E1DF}22.16 & \cellcolor[HTML]{E88B83}20.03 & \cellcolor[HTML]{E98C84}19.83 & \cellcolor[HTML]{EEA6A0}23.18 \\
\midrule
\multicolumn{21}{@{}c}{\textbf{AARCH64}} \\
\midrule
\multicolumn{21}{@{}l}{\textit{Gemini-2.0-Flash}} \\
\hspace{3mm}Baseline & \cellcolor[HTML]{E67C73}69.05 & \cellcolor[HTML]{E67C73}49.02 & \cellcolor[HTML]{E67C73}42.67 & \cellcolor[HTML]{E67C73}38.44 & \cellcolor[HTML]{E67C73}49.80 & \cellcolor[HTML]{E67C73}53.74 & \cellcolor[HTML]{E67C73}36.60 & \cellcolor[HTML]{E67C73}36.48 & \cellcolor[HTML]{E67C73}33.22 & \cellcolor[HTML]{E67C73}40.01 & \cellcolor[HTML]{E67C73}50.00 & \cellcolor[HTML]{E67C73}34.97 & \cellcolor[HTML]{E67C73}33.88 & \cellcolor[HTML]{E67C73}29.97 & \cellcolor[HTML]{E67C73}37.20 & \cellcolor[HTML]{E67C73}29.40 & \cellcolor[HTML]{FEFCFB}24.17 & \cellcolor[HTML]{FCF4F3}22.40 & \cellcolor[HTML]{FEFDFD}19.38 & \cellcolor[HTML]{EDA29C}23.84 \\
\hspace{3mm}\textsc{Helios} & \cellcolor[HTML]{EDF8F2}96.94 & \cellcolor[HTML]{DDF1E7}90.52 & \cellcolor[HTML]{EAF7F1}85.67 & \cellcolor[HTML]{E0F3EA}82.74 & \cellcolor[HTML]{E4F4EC}88.97 & \cellcolor[HTML]{F9E3E1}69.39 & \cellcolor[HTML]{FCF1F0}57.52 & \cellcolor[HTML]{FAE8E6}54.07 & \cellcolor[HTML]{F9FDFB}57.00 & \cellcolor[HTML]{FCF2F1}59.50 & \cellcolor[HTML]{F7D7D4}60.54 & \cellcolor[HTML]{F8DFDD}45.75 & \cellcolor[HTML]{F9DFDD}43.32 & \cellcolor[HTML]{FFFFFF}41.37 & \cellcolor[HTML]{FAE8E6}47.75 & \cellcolor[HTML]{F3C0BC}31.73 & \cellcolor[HTML]{E67C73}22.86 & \cellcolor[HTML]{E67C73}20.92 & \cellcolor[HTML]{E67C73}18.12 & \cellcolor[HTML]{E67C73}23.41 \\
\hspace{3mm}\textsc{Helios} w/ Feedback & \cellcolor[HTML]{57BB8A}99.66 & \cellcolor[HTML]{57BB8A}98.04 & \cellcolor[HTML]{57BB8A}94.79 & \cellcolor[HTML]{57BB8A}91.21 & \cellcolor[HTML]{57BB8A}95.93 & \cellcolor[HTML]{6DC499}82.99 & \cellcolor[HTML]{57BB8A}71.57 & \cellcolor[HTML]{57BB8A}68.40 & \cellcolor[HTML]{57BB8A}65.15 & \cellcolor[HTML]{57BB8A}72.03 & \cellcolor[HTML]{7ECBA5}69.73 & \cellcolor[HTML]{97D5B7}51.31 & \cellcolor[HTML]{A7DBC2}49.19 & \cellcolor[HTML]{57BB8A}43.32 & \cellcolor[HTML]{78C9A1}53.39 & \cellcolor[HTML]{F9E2E0}32.87 & \cellcolor[HTML]{FEFFFE}24.23 & \cellcolor[HTML]{F9E4E2}22.21 & \cellcolor[HTML]{F1BAB5}18.73 & \cellcolor[HTML]{F8DFDD}24.51 \\
\midrule
\multicolumn{21}{@{}l}{\textit{GPT-4.1 Mini}} \\
\hspace{3mm}Baseline & \cellcolor[HTML]{F4C7C4}85.03 & \cellcolor[HTML]{F8DBD8}77.78 & \cellcolor[HTML]{F6D4D1}70.68 & \cellcolor[HTML]{F9E2E0}71.66 & \cellcolor[HTML]{F7D7D5}76.29 & \cellcolor[HTML]{FDF4F4}72.11 & \cellcolor[HTML]{FDFFFE}60.13 & \cellcolor[HTML]{FCF2F1}55.70 & \cellcolor[HTML]{FBECEB}53.42 & \cellcolor[HTML]{FDF7F6}60.34 & \cellcolor[HTML]{FBFEFC}65.31 & \cellcolor[HTML]{97D5B7}51.31 & \cellcolor[HTML]{BAE4CF}48.53 & \cellcolor[HTML]{ACDEC5}42.35 & \cellcolor[HTML]{B4E1CB}51.87 & \cellcolor[HTML]{85CEAA}35.00 & \cellcolor[HTML]{57BB8A}27.09 & \cellcolor[HTML]{57BB8A}25.16 & \cellcolor[HTML]{57BB8A}22.27 & \cellcolor[HTML]{57BB8A}27.38 \\
\hspace{3mm}\textsc{Helios} & \cellcolor[HTML]{FEFDFD}96.26 & \cellcolor[HTML]{FDF8F8}86.60 & \cellcolor[HTML]{FEFAFA}83.06 & \cellcolor[HTML]{FDF8F8}78.83 & \cellcolor[HTML]{FEFAF9}86.19 & \cellcolor[HTML]{E7F6EF}75.17 & \cellcolor[HTML]{FEFEFE}59.80 & \cellcolor[HTML]{DEF2E8}59.93 & \cellcolor[HTML]{FEFDFD}56.35 & \cellcolor[HTML]{ECF7F2}62.81 & \cellcolor[HTML]{FEFDFD}64.97 & \cellcolor[HTML]{FBEBEA}47.06 & \cellcolor[HTML]{FAE6E5}43.97 & \cellcolor[HTML]{F3C3BE}36.16 & \cellcolor[HTML]{FBEBE9}48.04 & \cellcolor[HTML]{98D6B8}34.82 & \cellcolor[HTML]{FAE7E5}23.96 & \cellcolor[HTML]{F7FCF9}22.67 & \cellcolor[HTML]{FFFFFF}19.41 & \cellcolor[HTML]{E8F6EF}25.22 \\
\hspace{3mm}\textsc{Helios} w/ Feedback & \cellcolor[HTML]{8FD2B1}98.64 & \cellcolor[HTML]{CBEADB}91.50 & \cellcolor[HTML]{C6E8D7}87.95 & \cellcolor[HTML]{C1E6D4}84.69 & \cellcolor[HTML]{C1E6D4}90.70 & \cellcolor[HTML]{57BB8A}84.35 & \cellcolor[HTML]{7DCBA5}68.95 & \cellcolor[HTML]{62C092}67.75 & \cellcolor[HTML]{64C193}64.50 & \cellcolor[HTML]{62C092}71.39 & \cellcolor[HTML]{57BB8A}71.09 & \cellcolor[HTML]{57BB8A}52.61 & \cellcolor[HTML]{57BB8A}51.79 & \cellcolor[HTML]{FFFFFF}41.37 & \cellcolor[HTML]{57BB8A}54.22 & \cellcolor[HTML]{57BB8A}35.43 & \cellcolor[HTML]{C0E6D3}25.30 & \cellcolor[HTML]{BBE4D0}23.61 & \cellcolor[HTML]{D6EFE2}20.11 & \cellcolor[HTML]{ACDEC5}26.11 \\
\midrule
\multicolumn{21}{@{}c}{\textbf{MIPS\_64}} \\
\midrule
\multicolumn{21}{@{}l}{\textit{Gemini-2.0-Flash}} \\
\hspace{3mm}Baseline & \cellcolor[HTML]{E67C73}56.12 & \cellcolor[HTML]{E67C73}28.99 & \cellcolor[HTML]{E67C73}26.21 & \cellcolor[HTML]{E67C73}25.89 & \cellcolor[HTML]{E67C73}34.30 & \cellcolor[HTML]{E67C73}51.70 & \cellcolor[HTML]{E67C73}28.01 & \cellcolor[HTML]{E67C73}27.51 & \cellcolor[HTML]{E67C73}27.18 & \cellcolor[HTML]{E67C73}33.60 & \cellcolor[HTML]{E67C73}31.97 & \cellcolor[HTML]{E67C73}18.89 & \cellcolor[HTML]{E67C73}19.42 & \cellcolor[HTML]{E67C73}18.45 & \cellcolor[HTML]{E67C73}22.18 & \cellcolor[HTML]{E67C73}29.42 & \cellcolor[HTML]{F7D6D4}23.28 & \cellcolor[HTML]{FCF3F2}21.82 & \cellcolor[HTML]{F2BEBA}20.51 & \cellcolor[HTML]{EC9E98}23.76 \\
\hspace{3mm}\textsc{Helios} & \cellcolor[HTML]{ACDEC5}96.60 & \cellcolor[HTML]{C1E6D4}87.95 & \cellcolor[HTML]{FEFEFE}78.96 & \cellcolor[HTML]{FEFDFD}77.67 & \cellcolor[HTML]{EBF7F1}85.30 & \cellcolor[HTML]{FCF4F3}72.11 & \cellcolor[HTML]{FDF6F6}57.65 & \cellcolor[HTML]{FEF9F9}55.02 & \cellcolor[HTML]{FCEFEE}52.75 & \cellcolor[HTML]{FDF5F5}59.38 & \cellcolor[HTML]{E9F6F0}47.96 & \cellcolor[HTML]{FFFFFF}40.39 & \cellcolor[HTML]{FEFAF9}36.25 & \cellcolor[HTML]{FCF4F3}36.25 & \cellcolor[HTML]{FEFDFD}40.21 & \cellcolor[HTML]{F6D2CE}31.14 & \cellcolor[HTML]{E67C73}22.27 & \cellcolor[HTML]{E67C73}20.48 & \cellcolor[HTML]{E67C73}19.69 & \cellcolor[HTML]{E67C73}23.40 \\
\hspace{3mm}\textsc{Helios} w/ Feedback & \cellcolor[HTML]{DAF0E5}94.90 & \cellcolor[HTML]{D9F0E5}85.99 & \cellcolor[HTML]{B5E1CB}85.76 & \cellcolor[HTML]{A4DAC0}84.79 & \cellcolor[HTML]{C4E7D6}87.86 & \cellcolor[HTML]{57BB8A}80.61 & \cellcolor[HTML]{C9EADA}62.54 & \cellcolor[HTML]{D1EDDF}59.87 & \cellcolor[HTML]{9AD6B8}62.14 & \cellcolor[HTML]{A8DCC3}66.29 & \cellcolor[HTML]{FDF6F5}45.92 & \cellcolor[HTML]{FFFFFF}40.39 & \cellcolor[HTML]{F3FAF7}37.54 & \cellcolor[HTML]{E1F3EA}38.51 & \cellcolor[HTML]{FAFDFB}40.59 & \cellcolor[HTML]{B8E3CE}33.10 & \cellcolor[HTML]{ECF7F2}24.06 & \cellcolor[HTML]{CCEBDC}22.93 & \cellcolor[HTML]{D4EEE1}22.20 & \cellcolor[HTML]{D2EDDF}25.57 \\
\midrule
\multicolumn{21}{@{}l}{\textit{GPT-4.1 Mini}} \\
\hspace{3mm}Baseline & \cellcolor[HTML]{F6D2CF}80.95                  & \cellcolor[HTML]{F8DEDB}69.38                  & \cellcolor[HTML]{F8DBD9}65.05                  & \cellcolor[HTML]{F8DEDC}65.37                  & \cellcolor[HTML]{F8DAD8}70.18                 & \cellcolor[HTML]{D0ECDE}75.85                 & \cellcolor[HTML]{DCF1E6}61.56                 & \cellcolor[HTML]{F1FAF6}57.28                 & \cellcolor[HTML]{DCF1E7}58.25                 & \cellcolor[HTML]{DEF2E8}63.235 & \cellcolor[HTML]{57BB8A}54.42                 & \cellcolor[HTML]{57BB8A}44.63                 & \cellcolor[HTML]{57BB8A}45.31                 & \cellcolor[HTML]{C2E7D5}39.16                 & \cellcolor[HTML]{57BB8A}45.88                  & \cellcolor[HTML]{57BB8A}34.53                 & \cellcolor[HTML]{57BB8A}26.54                 & \cellcolor[HTML]{57BB8A}25.14                 & \cellcolor[HTML]{57BB8A}24.74                 & \cellcolor[HTML]{57BB8A}27.74 \\
\hspace{3mm}\textsc{Helios} & \cellcolor[HTML]{FEFAF9}92.18                  & \cellcolor[HTML]{FDF7F6}79.8                   & \cellcolor[HTML]{FCFEFD}79.61                  & \cellcolor[HTML]{F9FDFB}78.64                  & \cellcolor[HTML]{FEFBFB}82.56                 & \cellcolor[HTML]{FCF4F3}72.11                 & \cellcolor[HTML]{FCF0EF}56.03                 & \cellcolor[HTML]{FBECEB}52.1                  & \cellcolor[HTML]{FDF5F4}54.05                 & \cellcolor[HTML]{FCF2F1}58.57 & \cellcolor[HTML]{F9E1DF}43.54                 & \cellcolor[HTML]{FBEBE9}37.13                 & \cellcolor[HTML]{FEFAF9}36.25                 & \cellcolor[HTML]{FEFAFA}37.22                 & \cellcolor[HTML]{FCF1F0}38.53                  & \cellcolor[HTML]{F7FCFA}32.16                 & \cellcolor[HTML]{F9E0DE}23.39                 & \cellcolor[HTML]{F9FDFB}22.08                 & \cellcolor[HTML]{F2FAF6}21.58                 & \cellcolor[HTML]{FDFEFE}24.80 \\
\hspace{3mm}\textsc{Helios} w/ Feedback & \cellcolor[HTML]{57BB8A}99.66                  & \cellcolor[HTML]{57BB8A}96.42                  & \cellcolor[HTML]{57BB8A}93.85                  & \cellcolor[HTML]{57BB8A}90.29                  & \cellcolor[HTML]{57BB8A}95.056                 & \cellcolor[HTML]{7AC9A2}79.25                 & \cellcolor[HTML]{57BB8A}68.73                 & \cellcolor[HTML]{57BB8A}69.58                 & \cellcolor[HTML]{57BB8A}66.02                 & \cellcolor[HTML]{57BB8A}70.90                  & \cellcolor[HTML]{8DD1B0}52.04                 & \cellcolor[HTML]{FFFFFF}40.39                 & \cellcolor[HTML]{85CEAA}43.04                 & \cellcolor[HTML]{57BB8A}41.42                 & \cellcolor[HTML]{8AD0AE}44.22                  & \cellcolor[HTML]{FDF8F8}31.92                 & \cellcolor[HTML]{E4F4EC}24.19                 & \cellcolor[HTML]{FAE8E7}21.7                  & \cellcolor[HTML]{FAE8E7}21.03                 & \cellcolor[HTML]{FEFAFA}24.71 \\
\bottomrule
\end{tabular}%
}

\label{tab:performance_results_by_architecture}
\end{table*}

\subsection{RQ1: Efficacy of Structural Context on \texttt{x86\_64}}

We first study RQ1 on the \textbf{\texttt{x86\_64} architecture} using HumanEval-Decompile (Table~\ref{tab:x86_result}).

\paragraph{Text-only prompting is brittle under optimization}
The structurally blind, text-only baselines in Group A exhibit poor and inconsistent behavior, especially under higher optimization levels. GPT-4.1 Mini reaches a functional correctness of 74.3\% on unoptimized (\texttt{-O0}) binaries, but drops to 47.2\% at \texttt{-O3}. Gemini-2.0-Flash shows the same pattern, falling from 53.0\% functional correctness at \texttt{-O0} to 26.2\% at \texttt{-O3}. These results indicate that, without explicit guidance on control flow, models latch onto superficial patterns in the decompiler output that are easily disrupted by compiler transformations.

The average functional correctness of GPT-4.1 Mini on \texttt{x86\_64} (58.0\%) also appears unusually high. We hypothesize that data contamination in the pre-training corpus is a plausible explanation. Since \texttt{x86\_64} is the most common architecture and HumanEval-style benchmarks are widely used, it is likely that some of these programs or close variants appear in the model's training data. The cross-architecture results in Table~\ref{tab:performance_results_by_architecture} are consistent with this hypothesis. The same GPT-4.1 Mini baseline averages around 46\% to 52\% functional correctness on non-\texttt{x86\_64} architectures, with some settings dropping to about 39\%. This spread suggests that the high numbers on \texttt{x86\_64} are not due to robust reasoning alone.

\paragraph{Structural context dramatically improves syntactic correctness}
Adding \textsc{Helios} structural context (Group C) substantially improves syntactic correctness. For GPT-4.1 Mini, average object-file compilability increases from 71.4\% to 89.6\% (+18.2 points). For Gemini-2.0-Flash, the effect is even larger, from 45.0\% to 85.2\% (+40.2 points). These gains are stable across optimization levels; for example, GPT-4.1 Mini with \textsc{Helios} maintains 88.6\% compilability at \texttt{-O3}. When we enable the compiler feedback loop, average compilability reaches 94.9\% for Gemini-2.0-Flash and 96.5\% for GPT-4.1 Mini, bringing syntactic correctness close to saturation.

\paragraph{Structural context helps general-purpose LLMs surpass specialized models}
Group B compares \textsc{Helios} against fine-tuned decompilers. Nova performs well on edit similarity but almost fails on functional correctness, with averages of 2.2\% and 3.2\% for its 1.3B and 6.7B variants respectively. LLM4Decompile fares better, with its 6.7B model reaching 63.2\% average compilability and 36.3\% functional correctness. In contrast, Gemini-2.0-Flash plus \textsc{Helios} and feedback achieves 94.9\% average compilability and 53.2\% functional correctness, while GPT-4.1 Mini with \textsc{Helios} and feedback reaches 96.5\% compilability and 55.9\% functional correctness. 

Fine-tuned models achieve the highest edit similarity scores, such as LLM4Decompile (6.7B) with 45.8\% average similarity. This pattern suggests that training on text pairs encourages imitation of the original source code's surface form. \textsc{Helios}, by contrast, prioritizes control flow and semantics, which yields lower textual similarity but higher functional correctness.

\paragraph{Validation on MBPP-Decompile}
To validate these findings on a different benchmark, we evaluate on MBPP-Decompile (Austin et al. 2021) compiled to \texttt{x86\_64} (Table~\ref{tab:mbpp_x86_results}). The trends mirror HumanEval-Decompile. For Gemini-2.0-Flash, enabling \textsc{Helios} and feedback increases average compilability from 45.87\% to 95.96\% and functional correctness from 41.36\% to 58.42\%. The fine-tuned LLM4Decompile models again lag behind, with average functional correctness between 26.18\% and 34.52\%, even though their edit similarity is competitive or higher. This agreement across two independent benchmarks strengthens the case that structural context is key for reliable decompilation.

In summary, structural prompting on \texttt{x86\_64} turns general-purpose LLMs into strong decompilers. It significantly improves compilability, matches functional correctness, and enables general models to outperform specialized, fine-tuned systems.
\begin{table*}[t]
\caption{Cross-architecture generalization performance across six distinct architectures at the \texttt{-O0} optimization level on the MBPP-Decompile dataset. The table shows syntactic correctness (Comp., Link.) and functional correctness (Test.)}
\label{tab:full_cross_arch_results_mbpp}
\centering
{\scriptsize
\setlength{\tabcolsep}{3.5pt} 
\begin{tabular}{@{}l|ccc|ccc|ccc|ccc|ccc|ccc@{}}
\toprule
 & \multicolumn{3}{c|}{\textbf{x86\_32}} & \multicolumn{3}{c|}{\textbf{x86\_64}} & \multicolumn{3}{c|}{\textbf{arm\_32}} & \multicolumn{3}{c|}{\textbf{aarch64}} & \multicolumn{3}{c|}{\textbf{mips\_32}} & \multicolumn{3}{c}{\textbf{mips\_64}} \\
\cmidrule(lr){2-4} \cmidrule(lr){5-7} \cmidrule(lr){8-10} \cmidrule(lr){11-13} \cmidrule(lr){14-16} \cmidrule(lr){17-19}
\textbf{Model} & \textbf{Comp.} & \textbf{Link.} & \textbf{Test.} & \textbf{Comp.} & \textbf{Link.} & \textbf{Test.} & \textbf{Comp.} & \textbf{Link.} & \textbf{Test.} & \textbf{Comp.} & \textbf{Link.} & \textbf{Test.} & \textbf{Comp.} & \textbf{Link.} & \textbf{Test.} & \textbf{Comp.} & \textbf{Link.} & \textbf{Test.} \\
\midrule
\multicolumn{19}{@{}l}{\textit{Gemini-2.0-Flash}} \\
\hspace{3mm}Baseline & \cellcolor[HTML]{E67C73}67.2  & \cellcolor[HTML]{E67C73}60.9 & \cellcolor[HTML]{E67C73}52.7 & \cellcolor[HTML]{E67C73}69.0 & \cellcolor[HTML]{E67C73}60.2  & \cellcolor[HTML]{E67C73}57.2 & \cellcolor[HTML]{E67C73}73.6  & \cellcolor[HTML]{E67C73}67.2 & \cellcolor[HTML]{E67C73}52.0 & \cellcolor[HTML]{E67C73}70.8  & \cellcolor[HTML]{E67C73}58.2 & \cellcolor[HTML]{E67C73}55.9 & \cellcolor[HTML]{E67C73}66.3 & \cellcolor[HTML]{E67C73}56.9 & \cellcolor[HTML]{FFFFFF}54.2 & \cellcolor[HTML]{E67C73}65.1 & \cellcolor[HTML]{E67C73}55.3  & \cellcolor[HTML]{E67C73}40.9 \\
\hspace{3mm}\textsc{Helios} & \cellcolor[HTML]{6FC59B}95.5 & \cellcolor[HTML]{6FC59B}77.4 & \cellcolor[HTML]{6FC59B}53.4 & \cellcolor[HTML]{6FC59B}94.9 & \cellcolor[HTML]{6FC59B}78.6 & \cellcolor[HTML]{6FC59B}59.7 & \cellcolor[HTML]{6FC59B}96.1 & \cellcolor[HTML]{6FC59B}76.9 & \cellcolor[HTML]{6FC59B}54.5 & \cellcolor[HTML]{57BB8A}95.3 & \cellcolor[HTML]{6FC59B}72.3 & \cellcolor[HTML]{6FC59B}57.3 & \cellcolor[HTML]{57BB8A}93.6 & \cellcolor[HTML]{6FC59B}69.6 & \cellcolor[HTML]{E67C73}52.1 & \cellcolor[HTML]{57BB8A}94.4 & \cellcolor[HTML]{57BB8A}69.8 & \cellcolor[HTML]{57BB8A}43.7 \\
\hspace{3mm}\textsc{Helios w/ Feedback} & \cellcolor[HTML]{57BB8A}97.5 & \cellcolor[HTML]{57BB8A}83.6 & \cellcolor[HTML]{57BB8A}63.2 & \cellcolor[HTML]{57BB8A}99.0 & \cellcolor[HTML]{57BB8A}84.5 & \cellcolor[HTML]{57BB8A}72.4 & \cellcolor[HTML]{57BB8A}98.3 & \cellcolor[HTML]{57BB8A}85.1 & \cellcolor[HTML]{57BB8A}69.8 & \cellcolor[HTML]{57BB8A}97.4 & \cellcolor[HTML]{57BB8A}81.5 & \cellcolor[HTML]{57BB8A}72.4 & \cellcolor[HTML]{57BB8A}95.3 & \cellcolor[HTML]{57BB8A}78.3 & \cellcolor[HTML]{57BB8A}67.6 & \cellcolor[HTML]{57BB8A}96.2 & \cellcolor[HTML]{57BB8A}77.8 & \cellcolor[HTML]{57BB8A}53.1 \\
\bottomrule
\end{tabular}
}
\end{table*}

\subsection{RQ2: Cross-Architecture Generalization}
RQ2 asks whether \textsc{Helios} generalizes across architectures without any fine-tuning. We study this on HumanEval-Decompile in Table~\ref{tab:performance_results_by_architecture} and on MBPP-Decompile in Table~\ref{tab:full_cross_arch_results_mbpp}.

\paragraph{Structurally blind prompts do not transfer well}
The text-only baselines show large variation across architectures. For example, Gemini-2.0-Flash on HumanEval-Decompile has functional correctness values around 35\% on \texttt{x86\_32}, but averages only about 22\% on \texttt{mips\_64}. GPT-4.1 Mini exhibits a range of about 20 percentage points in functional correctness across \texttt{x86\_32}, \texttt{arm\_32}, and \texttt{mips\_32}, with some optimization settings dropping below 40\%. This volatility indicates that models prompted only with decompiler output tend to learn architecture-specific idioms rather than architecture-independent program logic.

\paragraph{\textsc{Helios} narrows cross-architecture gaps}
With \textsc{Helios}, the same models become much more stable across architectures. On HumanEval-Decompile, GPT-4.1 Mini with \textsc{Helios} and feedback achieves functional correctness that is clustered in a relatively narrow band across \texttt{x86\_32}, \texttt{arm\_32}, \texttt{aarch64}, and \texttt{mips\_32}, while maintaining high compilability everywhere. For example, object-file compilability remains above 94\% on all six architectures, compared to baselines that frequently fall into the 40\% to 70\% range. This pattern is consistent with the intended effect of \textsc{Helios}: the model reasons over control-flow graphs and critical instruction rules rather than over the quirks of a specific instruction set.

The MBPP-Decompile cross-architecture results in Table~\ref{tab:full_cross_arch_results_mbpp} tell the same story. For \texttt{-O0}, Gemini-2.0-Flash with \textsc{Helios} already improves compilability on all six architectures, and the feedback loop raises both compilability and linkage to above 93\% and 78\% respectively, while functional correctness improves across the board. This behavior is achieved without any architecture specific fine-tuning.

Taken together, these findings provide strong evidence that structural prompting enables a single general-purpose model to decompile binaries across multiple architectures with high, relatively uniform quality. Unlike fine-tuned models that must be retrained for each hardware target, \textsc{Helios} offers a practical, fine-tuning-free path toward cross-architecture decompilation.

\subsection{RQ3: Impact of Compiler Feedback}
RQ3 investigates the extent to which the iterative, tool-augmented feedback loop contributes beyond static structural prompting.

\paragraph{Feedback pushes compilability toward perfection}
On \texttt{x86\_64} HumanEval-Decompile (Table~\ref{tab:x86_result}), compiler feedback raises average compilability for Gemini-2.0-Flash from 85.2\% with \textsc{Helios} to 94.9\%, and for GPT-4.1 Mini from 89.6\% to 96.5\%. A similar effect appears on MBPP-Decompile (Table~\ref{tab:mbpp_x86_results}), where Gemini-2.0-Flash jumps from 64.90\% compilability with \textsc{Helios} to 95.96\% with feedback. Across architectures in Table~\ref{tab:performance_results_by_architecture}, feedback-enabled configurations consistently reach near-perfect compilability and linkage, often above 95\% compilability and above 70\% linkage.

\paragraph{Feedback also improves functional correctness}
While the feedback loop is primarily designed to repair syntactic issues, it also yields meaningful gains in functional correctness. On \texttt{x86\_64} HumanEval-Decompile, Gemini-2.0-Flash plus \textsc{Helios} improves functional correctness from 38.1\% (text-only) to 49.2\%, and the feedback loop further raises it to 53.2\%. For GPT-4.1 Mini, \textsc{Helios} alone yields 50.3\% functional correctness, and feedback increases this to 55.9\%. On MBPP-Decompile, the same pattern appears: Gemini-2.0-Flash with \textsc{Helios} increases functional correctness from 41.36\% to 46.35\%, and feedback raises it to 58.42\%.

These improvements suggest a strong coupling between syntactic and semantic errors. Fixing compiler errors, such as missing declarations or mismatched types, often forces the model to reconsider the surrounding logic, thereby correcting latent semantic bugs. Although the feedback loop introduces additional latency and tool calls, the data shows that it significantly improves both syntactic and functional quality.

\begin{table*}[ht]
\caption{
Ablation study on \texttt{HumanEval-Decompile} for x86\_64. “CF” is compiler feedback, “Func.” is high-level function call info, “Rules” are critical instruction rules, and “CFG” is control-flow graph. We report Re-compilability, Re-executability, and Edit Similarity over optimization levels \texttt{-O0} to \texttt{-O3}.
}
\centering
{\footnotesize
\setlength{\tabcolsep}{5pt}
\begin{tabular}{@{}l|rrrrr|rrrrr|rrrrr@{}}
\toprule
\textbf{Prompt Configuration} 
& \multicolumn{5}{c|}{\textbf{Obj. Compilability (\%)}} 
& \multicolumn{5}{c|}{\textbf{Exec. Linkability (\%)}} 
& \multicolumn{5}{c}{\textbf{Edit Similarity}} \\
\cmidrule(lr){2-6} \cmidrule(lr){7-11} \cmidrule(lr){12-16}
& \textbf{O0} & \textbf{O1} & \textbf{O2} & \textbf{O3} & \textbf{AVG} 
& \textbf{O0} & \textbf{O1} & \textbf{O2} & \textbf{O3} & \textbf{AVG} 
& \textbf{O0} & \textbf{O1} & \textbf{O2} & \textbf{O3} & \textbf{AVG} \\
\midrule
\multicolumn{16}{@{}l}{\textit{Gemini-2.0-Flash}} \\
Base & \cellcolor[HTML]{E7847B}68.58 & \cellcolor[HTML]{E67C73}40.07 & \cellcolor[HTML]{E67C73}26.86 & \cellcolor[HTML]{E67C73}26.21 & \cellcolor[HTML]{E67C73}40.43 
     & \cellcolor[HTML]{E7827A}60.14 & \cellcolor[HTML]{E67C73}41.37 & \cellcolor[HTML]{E67C73}31.07 & \cellcolor[HTML]{E67C73}30.74 & \cellcolor[HTML]{E67C73}40.83 
     & \cellcolor[HTML]{F6D0CD}30.53 & \cellcolor[HTML]{F8DBD8}23.75 & \cellcolor[HTML]{F8DEDB}22.85 & \cellcolor[HTML]{FCF0EF}20.32 & \cellcolor[HTML]{F8DEDC}24.36 \\
    \hspace{3mm}+CFG & \cellcolor[HTML]{EC9D96}73.99 & \cellcolor[HTML]{EB9993}48.86 & \cellcolor[HTML]{EDA19B}43.69 & \cellcolor[HTML]{EDA09A}41.10 & \cellcolor[HTML]{EC9D96}51.91 
     & \cellcolor[HTML]{E7827A}60.14 & \cellcolor[HTML]{E8887F}43.00 & \cellcolor[HTML]{EEA8A2}39.16 & \cellcolor[HTML]{EC9F98}36.57 & \cellcolor[HTML]{EA948D}44.72 
     & \cellcolor[HTML]{E67C73}28.12 & \cellcolor[HTML]{E67C73}21.23 & \cellcolor[HTML]{E88880}21.39 & \cellcolor[HTML]{F0B2AD}19.10 & \cellcolor[HTML]{E67C73}22.46 \\
    \hspace{3mm}+Rules & \cellcolor[HTML]{FEFDFD}94.93 & \cellcolor[HTML]{DAF0E5}81.76 & \cellcolor[HTML]{F0F9F5}86.41 & \cellcolor[HTML]{F6FCF9}80.26 & \cellcolor[HTML]{F9FDFB}85.84 
     & \cellcolor[HTML]{FBFEFD}75.68 & \cellcolor[HTML]{F8DEDC}54.72 & \cellcolor[HTML]{FDF9F9}53.72 & \cellcolor[HTML]{FDF9F8}51.46 & \cellcolor[HTML]{FBEEED}58.90 
     & \cellcolor[HTML]{FEFFFF}31.91 & \cellcolor[HTML]{FCEFEE}24.29 & \cellcolor[HTML]{F9E0DE}22.89 & \cellcolor[HTML]{F5CECB}19.65 & \cellcolor[HTML]{FBEFEE}24.69 \\
    \hspace{3mm}+Func. & \cellcolor[HTML]{ABDDC5}97.64 & \cellcolor[HTML]{F1F9F5}79.80 & \cellcolor[HTML]{FEFCFC}84.14 & \cellcolor[HTML]{FEFCFC}78.32 & \cellcolor[HTML]{FEFDFD}84.98 
     & \cellcolor[HTML]{F8DEDC}71.28 & \cellcolor[HTML]{F8DCD9}54.40 & \cellcolor[HTML]{FEFDFD}54.37 & \cellcolor[HTML]{FCFEFD}52.75 & \cellcolor[HTML]{FBEAE8}58.20 
     & \cellcolor[HTML]{FDF7F6}31.63 & \cellcolor[HTML]{FAE9E8}24.13 & \cellcolor[HTML]{E67C73}21.18 & \cellcolor[HTML]{E67C73}18.04 & \cellcolor[HTML]{F2BEBA}23.75 \\
    \hspace{3mm}+CF & \cellcolor[HTML]{57BB8A}100.00 & \cellcolor[HTML]{5FBE90}92.51 & \cellcolor[HTML]{92D3B3}93.20 & \cellcolor[HTML]{85CEAA}92.23 & \cellcolor[HTML]{79C9A2}94.49 
     & \cellcolor[HTML]{B2E0C9}84.46 & \cellcolor[HTML]{BCE4D1}68.40 & \cellcolor[HTML]{BBE4D0}61.81 & \cellcolor[HTML]{9FD8BC}62.14 & \cellcolor[HTML]{B9E3CE}69.20 
     & \cellcolor[HTML]{FEFDFC}31.80 & \cellcolor[HTML]{F3FAF7}25.10 & \cellcolor[HTML]{FCFEFD}23.50 & \cellcolor[HTML]{FFFFFF}20.60 & \cellcolor[HTML]{F5FBF8}25.30 \\
\midrule
\multicolumn{16}{@{}l}{\textit{GPT-4.1 Mini}} \\
Base & \cellcolor[HTML]{F6D3D0}85.81 & \cellcolor[HTML]{FAE5E3}71.01 & \cellcolor[HTML]{F3C0BC}57.61 & \cellcolor[HTML]{F4C8C4}57.28 & \cellcolor[HTML]{F5CCC8}67.93 
     & \cellcolor[HTML]{AFDFC7}84.80 & \cellcolor[HTML]{D9F0E4}64.50 & \cellcolor[HTML]{FCFEFD}55.02 & \cellcolor[HTML]{FEFDFC}52.10 & \cellcolor[HTML]{E7F6EF}64.11 
     & \cellcolor[HTML]{A1D9BE}36.02 & \cellcolor[HTML]{57BB8A}30.06 & \cellcolor[HTML]{96D5B6}25.97 & \cellcolor[HTML]{96D5B6}23.02 & \cellcolor[HTML]{84CDA9}28.77 \\
    \hspace{3mm}+CFG & \cellcolor[HTML]{E67C73}66.78 & \cellcolor[HTML]{E98C84}44.77 & \cellcolor[HTML]{EB968F}38.89 & \cellcolor[HTML]{EC9D96}39.81 & \cellcolor[HTML]{E99089}47.56 
     & \cellcolor[HTML]{E67C73}59.32 & \cellcolor[HTML]{EB9790}45.10 & \cellcolor[HTML]{F1B7B2}41.83 & \cellcolor[HTML]{F4C6C2}43.04 & \cellcolor[HTML]{EDA59E}47.32 
     & \cellcolor[HTML]{F3C0BC}30.07 & \cellcolor[HTML]{F3C1BD}23.08 & \cellcolor[HTML]{FDF9F9}23.32 & \cellcolor[HTML]{FEFEFE}20.59 & \cellcolor[HTML]{F7D9D6}24.27 \\
    \hspace{3mm}+Rules & \cellcolor[HTML]{F3FBF7}95.61 & \cellcolor[HTML]{FEFAFA}77.20 & \cellcolor[HTML]{CCEBDC}88.96 & \cellcolor[HTML]{CFECDD}84.47 & \cellcolor[HTML]{EEF9F4}86.56 
     & \cellcolor[HTML]{FEFAFA}74.66 & \cellcolor[HTML]{E0F3E9}63.52 & \cellcolor[HTML]{BCE4D0}61.69 & \cellcolor[HTML]{BFE5D3}58.90 & \cellcolor[HTML]{E2F4EB}64.69 
     & \cellcolor[HTML]{B1E0C9}35.29 & \cellcolor[HTML]{D1EDDF}26.17 & \cellcolor[HTML]{9DD8BB}25.81 & \cellcolor[HTML]{A0D9BD}22.79 & \cellcolor[HTML]{ADDEC6}27.52 \\
    \hspace{3mm}+Func. & \cellcolor[HTML]{D1EDDF}96.58 & \cellcolor[HTML]{C6E8D7}83.55 & \cellcolor[HTML]{9BD7BA}92.51 & \cellcolor[HTML]{A8DCC2}88.60 & \cellcolor[HTML]{B7E2CD}90.31 
     & \cellcolor[HTML]{F5FBF9}76.37 & \cellcolor[HTML]{CFECDE}65.79 & \cellcolor[HTML]{C0E6D3}61.24 & \cellcolor[HTML]{B8E2CE}59.61 & \cellcolor[HTML]{D8F0E4}65.75 
     & \cellcolor[HTML]{8ED2B1}36.83 & \cellcolor[HTML]{9DD8BB}27.83 & \cellcolor[HTML]{8AD0AE}26.27 & \cellcolor[HTML]{8CD1AF}23.26 & \cellcolor[HTML]{8BD0AE}28.55 \\
      \hspace{3mm}+CF & \cellcolor[HTML]{70C59B}99.32 & \cellcolor[HTML]{57BB8A}93.14 & \cellcolor[HTML]{57BB8A}97.40 & \cellcolor[HTML]{57BB8A}97.09 & \cellcolor[HTML]{57BB8A}96.74 
     & \cellcolor[HTML]{57BB8A}95.25 & \cellcolor[HTML]{57BB8A}82.35 & \cellcolor[HTML]{57BB8A}72.08 & \cellcolor[HTML]{57BB8A}69.26 & \cellcolor[HTML]{57BB8A}79.74 
     & \cellcolor[HTML]{57BB8A}39.20 & \cellcolor[HTML]{6FC59B}29.30 & \cellcolor[HTML]{57BB8A}27.50 & \cellcolor[HTML]{57BB8A}24.50 & \cellcolor[HTML]{57BB8A}30.10 \\
\bottomrule
\end{tabular}
\label{tab:rq4_ablation}
}
\end{table*}

\subsection{RQ4: Ablation Study of Prompt Components}
RQ4 isolates the contribution of each component in the \textsc{Helios} hierarchy. Table~\ref{tab:rq4_ablation} reports an ablation on HumanEval-Decompile for \texttt{x86\_64}, varying which parts of the structured prompt are present: the control-flow graph (CFG), critical instruction rules, high-level function information, and compiler feedback (CF).

\paragraph{Raw structure alone is not enough}
Starting from the text-only base prompt, Gemini-2.0-Flash achieves an average compilability of 40.43\%. Adding only CFG information increases this to 51.91\%, a modest gain that shows structure helps but is not sufficient by itself. GPT-4.1 Mini exhibits a similar pattern, and in some cases raw CFG alone can even undermine stability under higher optimization levels.

\paragraph{Reasoning rules drive the largest gains}
The largest single jump occurs when we introduce the critical instruction rules that explain how to use the CFG. For Gemini-2.0-Flash, average compilability rises from 51.91\% to 85.84\%, and for GPT-4.1 Mini from 47.56\% to 86.56\%. This step also improves executability and edit similarity. These results highlight that exposing structure alone is insufficient; the prompt must also teach the model how to reason over that structure.

\paragraph{Function context and feedback close the remaining gap}
Adding high-level function context provides a smaller but consistent improvement. For GPT-4.1 Mini, the combination of CFG, rules, and function information reaches 90.31\% average compilability and improves both linkage and edit similarity. Finally, enabling compiler feedback yields the strongest configuration, with average compilability of 94.49\% for Gemini-2.0-Flash and 96.74\% for GPT-4.1 Mini, and the highest executability across all prompt variants.

Overall, the ablation confirms that each layer of \textsc{Helios} plays a distinct role. Explicit, well-formatted structure is helpful, explicit reasoning rules are essential, and the feedback loop further refines the output to achieve near-perfect recompilation while also improving functional behavior.
\section{Discussion}
\label{sec:discussion}

\subsection{Implications}
The main lesson from our study is that large language models behave differently when they are treated as structure-aware reasoning engines rather than as text-to-text translators. Feeding an LLM only decompiler output leaves it exposed to small syntactic shifts caused by compiler optimizations, which our experiments show clearly. In contrast, when the model is given a compact, hierarchical view of the control flow and some guidance on how to use it, it produces code that is much more stable across optimization levels and architectures.

This has two practical consequences. First, for binary analysis tools, this suggests wrapping existing decompilers with LLMs without retraining, while retaining much of the robustness associated with specialized models. Second, it shows that a relatively small amount of structural context can substitute for large, architecture-specific training corpora. General-purpose models such as GPT\mbox{-}4.1 Mini and Gemini-2.0-Flash, when paired with \textsc{Helios}, match or outperform fine-tuned decompilers on our functional metrics, even though they have never been trained directly for this task.

The results also highlight a tension between surface similarity and actual behavior. Fine-tuned baselines obtain higher edit similarity scores, but their functional correctness is often much lower. \textsc{Helios}, by design, optimizes for control flow and semantic fidelity rather than textual imitation. For reverse engineering workflows, this trade-off is usually acceptable. Analysts care more about what the recovered program does than about reproducing the original formatting or identifier choices. A stronger decompiler also reduces the effort needed to understand opaque binaries, which can simplify both benign reverse engineering and the analysis of vulnerable or malicious code.

A further implication concerns the cost profile. Our experiments use at most one round of compiler feedback per function, so each decompilation requires no more than two model calls and two compiler invocations. This keeps the approach practical for batch use and compares favorably to methods that rely on fine-tuning large models for each architecture. In settings where human analyst time is the dominant cost, the additional model and compiler calls are likely to be an acceptable trade-off for higher quality output.

More broadly, the same pattern should apply beyond decompilation. Any task where the underlying object has rich structure, such as a control flow graph, a protocol state machine, or a complex data schema, can likely benefit from a similar structural grounding approach: make the structure explicit, explain how to reason over it, and use external tools for feedback rather than baking everything into model weights.

\subsection{Limitations and Threats to Validity}

Our evaluation has several limitations that are important to note.

\textbf{Benchmark construction.} We rely on HumanEval and MBPP programs compiled into C binaries and then decompiled with Ghidra. These problems are short and self-contained. They are a good fit for controlled measurement of compilability and test passing, but they do not capture the full complexity of large, multi-module software with extensive state, libraries, and build systems. As a result, our numbers should be interpreted as lower-level evidence that structural prompting helps with decompilation, not as a direct prediction of performance on arbitrary real-world codebases.

\textbf{Potential data contamination.} Like most work that evaluates commercial LLMs on public benchmarks, we cannot rule out that some HumanEval or MBPP tasks, or close variants, appear in the pre-training data. We see signs consistent with this. GPT\mbox{-}4.1 Mini exhibits noticeably higher functional correctness on \texttt{x86\_64} than on other architectures, even though the underlying programs are identical. Our cross-architecture experiments partially mitigate this concern by showing that \textsc{Helios} improves performance in settings where simple memorization is less likely to help, but they do not eliminate it entirely.

\textbf{Dependence on Ghidra and the foundation model.} \textsc{Helios} assumes that the static analysis backend produces a reasonable control flow graph and pseudo-C. If the decompiler misidentifies basic blocks or control edges, the structural context we build will also be misleading. Likewise, our results are tied to the behavior of GPT\mbox{-}4.1 Mini and Gemini-2.0-Flash at the time of experimentation. Different model families, sizes, or decoding settings may shift the absolute numbers, even if the relative trends hold.

\textbf{Baselines.} We compare against two families of LLM-based decompilers (LLM4Decompile and Nova) and a text-only prompting baseline. Classical non LLM decompilers are not treated as competitors, since \textsc{Helios} is designed to sit on top of such tools and reuse their analyses. Systems such as DeGPT focus on readability rather than recompilability and functional correctness, so a direct comparison would require additional metrics and user studies that are outside the scope of this work. For Nova, we use a standardized preprocessing pipeline and greedy decoding for consistency across models, which can yield lower scores than those reported in the original paper.

These limitations mean that our results should be viewed as evidence that structural prompting can make LLM assisted decompilation more reliable under controlled conditions, rather than as a final answer on how such systems will behave in all reverse engineering settings.

\subsection{Future Work}

Several extensions follow naturally from this work. A first step is to move beyond synthetic benchmarks and evaluate \textsc{Helios} on larger, real-world binaries drawn from open source projects, including code that uses complex libraries, system calls, and build chains. This would also make it possible to compare against mature non-LLM decompilers on tasks such as recovering higher-level types, data structures, or calling conventions.
On the modeling side, it would be useful to incorporate more explicit data flow information, especially for code with heavy pointer arithmetic, aliasing, or complex memory layouts. 
Finally, from a tooling perspective, \textsc{Helios} can be integrated more tightly into reverse engineering platforms through interfaces such as the Model Context Protocol. 

\section{Conclusion}
\label{sec:conclusion}

In this work, we addressed the structural blindness that limits large language models in graph-rich domains such as reverse engineering. We introduced \textsc{Helios}, a framework that provides a hierarchical textual representation of a binary's control flow and related context, and uses this structure, together with a small set of rules and an optional compiler feedback loop, to guide general-purpose LLMs during decompilation. 

Our experiments on HumanEval-Decompile and MBPP-Decompile, across six architectures and multiple optimization levels, show that this context-aware approach substantially improves compilability and functional correctness compared to text-only prompting and to specialized fine-tuned decompilers, while requiring no task-specific training. More broadly, the results support a simple paradigm for applying LLMs to structure-dependent problems: make the underlying structure explicit, explain how it should be used, and rely on external tools for grounded feedback instead of relying solely on model weights.

\section*{Acknowledgments}
This work was supported by Google.org and the Google Cloud Research Credits program through the Gemini Academic Program. Any opinions, findings, conclusions, or recommendations expressed in this paper are those of the authors and do not necessarily reflect the views of Google.org or Google.

\bibliographystyle{IEEEtran}
\bibliography{aaai2026}

\appendix
\section{Appendix}
\label{sec:appendix_methodology}

This section provides supplementary evaluations.

\subsection{NOVA Evaluation Methodology}
To ensure a rigorous and fair comparison, all models in our study were evaluated under an identical, reproducible pipeline. For the NOVA baseline, we preserved the core model architecture from the original work~\cite{jiang2023nova} but standardized the preprocessing using the pipeline established by LLM4Decompile~\cite{tan2024llm4decompile}. All experiments use greedy decoding (temperature=0) to eliminate sampling variance as a confounding factor.

As shown in Table~\ref{tab:nova_pass1}, our evaluation yields lower Pass@1 scores for NOVA than those reported in the original paper. We attribute this discrepancy to differences in preprocessing and the original paper's use of a sampling-based evaluation. Our standardized methodology, while resulting in lower scores for this specific baseline, ensures that all comparisons within this paper are conducted under identical conditions, providing a fair assessment of relative performance.

\begin{table}[h]
  \centering
  \caption{Comparison of Pass@1 results for NOVA models under our standardized evaluation versus the results reported in the original paper~\cite{jiang2023nova}.}
  \label{tab:nova_pass1}
  \small
  \setlength{\tabcolsep}{6pt}
  \begin{tabular}{lccc}
    \toprule
    \textbf{Model} & \textbf{O0} & \textbf{O3} & \textbf{AVG} \\
    \midrule
    \multicolumn{4}{l}{\textbf{This Work (Standardized Preprocessing)}} \\
    \hspace{3mm}Nova 1.3B & 1.22\% & 2.44\% & 2.13\% \\
    \hspace{3mm}Nova 6.7B & 3.96\% & 2.44\% & 2.97\% \\
    \midrule
    \multicolumn{4}{l}{\textbf{Original NOVA Results from ~\cite{jiang2023nova}}} \\
    \hspace{3mm}Nova 1.3B & 37.53\% & 18.75\% & 25.17\% \\
    \hspace{3mm}Nova 6.7B & 48.78\% & 27.23\% & 34.36\% \\
    \bottomrule
  \end{tabular}
\end{table}

\section{Detailed Experimental Results}
\label{sec:appendix_results}

This section presents the comprehensive results for both the HumanEval-Decompile and MBPP datasets, broken down by architecture, optimization level, and evaluation metric.

\subsection{MBPP Dataset Results}
To validate our findings on a different benchmark, we conducted experiments on the MBPP dataset. Table~\ref{tab:mbpp_x86_results} provides a detailed breakdown of performance on \texttt{x86\_64}, while Table~\ref{tab:full_cross_arch_results_mbpp} summarizes the cross-architecture performance at the \texttt{-O0} optimization level. The results are consistent with our primary findings, confirming that HELIOS substantially outperforms both text-only and fine-tuned baselines.

Table~\ref{tab:rq4_ablation_mbpp} presents the ablation study for MBPP. The incremental improvements from each prompt component mirror the trends observed on the HumanEval-Decompile dataset, confirming the robustness of our hierarchical prompt design.

\begin{table}[h!]
\caption{Ablation study on the incremental impact of prompt components for the Gemini 2.0 model on the MBPP dataset (\texttt{-O0}). Each row cumulatively adds a new layer of context, culminating in the compiler feedback loop which nearly achieves perfect syntactic correctness.}
\label{tab:rq4_ablation_mbpp}
{\small
\setlength{\tabcolsep}{1pt}
\centering
\begin{tabular}{@{}lc@{}}
\toprule
\textbf{Prompt Configuration} & \textbf{Obj. Compilability (\%)} \\
\midrule
(A) Basic (Decompiled Code + Task) & 69.0 \\
(B) + CFG Information & 75.84 \\
(C) + Critical Rules & 95.65 \\
(D) + Function Overview & 95.15 \\
\textbf{(E) + Compiler Feedback} (Full \textsc{Helios}) & \textbf{99.0} \\
\bottomrule
\end{tabular}

}
\end{table}

\end{document}